\documentclass{article}
\usepackage{graphicx} % Required for inserting images
\usepackage{amssymb}
\usepackage{amsmath}
\usepackage{amsthm}
\usepackage{color}
\usepackage{comment}
\usepackage{enumitem}		% for fancy lists
\usepackage{geometry}		% allows easy margin 
\usepackage{graphicx}
\usepackage{authblk}
\usepackage{amssymb}
\usepackage{listings}
\usepackage{xcolor}
\usepackage{caption}
\usepackage[linesnumbered,ruled,vlined]{algorithm2e}
\usepackage{color}
\usepackage{algpseudocode}
\usepackage{subcaption}
\usepackage{hyperref}
\usepackage{url}
\newtheorem{theorem}{Theorem}
\title{Discrete Lorenz Attractors in 3D Sinusoidal Maps}
\author[a]{Sishu Shankar Muni}
\affil[a]{School of Digital Sciences, Digital University Kerala, Technopark phase-IV Campus, Mangalapuram, 695317, Thiruvananthapuram, India}
\usepackage[english]{babel}

\begin{document}

\maketitle
\begin{abstract}
Discrete Lorenz attractors can be found in three-dimensional discrete maps. Discrete Lorenz attractors have similar topology to that of the continuous Lorenz attractor exhibited by the well studied 3D Lorenz system. However, the routes to the formation of discrete Lorenz attractor in 3D maps is different from that of the routes of formation of continuous Lorenz flow attractor. This paper explores the exotic dynamics of a three-dimensional sinusoidal map, highlighting the existence of discrete Lorenz attractors. Through a detailed bifurcation analysis, we discuss various formation pathways involving supercritical and subcritical Neimark-Sacker bifurcations, and homoclinic butterflies towards the formation of discrete Lorenz attractors. Using a two-parameter Lyapunov chart analysis, we systematically illustrate chaotic regions and periodic regions in the parameter space, illustrating diverse attractor topologies, including hyperchaotic and wavy chaotic structures.  Furthermore, we investigate spatiotemporal patterns in a ring-star network of coupled sinusoidal maps, revealing diverse spatiotemporal patterns such as chimera states, traveling waves, and synchronization transitions, controlled by coupling strengths of ring and star networks respectively. A key contribution is the application of hyperchaotic signals to secure video encryption, where we apply YOLOv10 for object detection and CNN-based cryptographic key generation. This selective encryption approach ensures enhanced security, high computational efficiency, and robustness against differential attacks.
\end{abstract}

\section*{Introduction}
Discrete maps are remarkable and the simplest examples of systems exhibiting chaotic dynamics. One-dimensional maps can even show chaos \cite{May1976}. Since the beginning of the field of chaos, chaotic attractors, strange attractors caught a great attention \cite{Oestreicher2007}. Researchers started investigating one-dimensional and two-dimensional maps and have till now mostly understood the pathways of formation of chaotic attractors. However, scarce attention has been paid to the dynamics of multidimensional maps such as three-dimensional maps or more. Around 1990's, there were few studies attempting to study the dynamics of three-dimensional maps \cite{Robinson1998}. Gonchenko et. al. \cite{Gonchenko2014} showed various possible routes to chaotic dynamics in three-dimensional maps. Three-dimensional maps are important as they serve as an effective tool to study the four-dimensional flows. The study of multidimensional maps is interesting because they show additional new bifurcations which are absent in lower dimensional maps \cite{MuniHR}. They are the doubling bifurcation of ergodic and resonant torus. Three-dimensional maps exhibit the doubling bifurcation of quasiperiodic orbit.  Since 1980's, doubling bifurcation of quasiperiodic orbits was reported by Kaneko \cite{Kaneko83}. Since then, it was reported in various applied dynamical systems like Goodwins oscillator \cite{Zhu21}, radiophysical generators \cite{Kuznetsov2024}, neuron mappings \cite{MuniHR}, mechanical systems \cite{Li2022}. More recently, it was shown that doubling bifurcation of resonant torus or mode-locked orbit can also occur \cite{MuniTorus}. Various examples of 3D maps were shown which can exhibit both types of resonant torus doubling bifurcation such as a) length-doubled resonant torus, and b) disjoint length-doubled resonant torus. Moreover examples of 3D maps were given which shows both doubling bifurcation of quasiperiodic orbit and mode-locked orbit \cite{MuniTorus}. 

Another beautiful application of three-dimensional maps is that they showcase 
discrete Lorenz attractors, very similar in structure to that of the Lorenz attractor in 3D Lorenz system. Lorenz attractor was found by E. Lorenz in 1963 \cite{Lorenz1963}. Since then, it has led to many new research directions. It has led to the theory of singular hyperbolic attractors \cite{Arajo2021}, and also to the theory of pseudohyperbolic attractors \cite{Kainov2021}. Lorenz attractors are formed after a pitchfork bifurcation of an equilibrium point. Lorenz attractor is singularly hyperbolic that is it does not admit any homoclinic tangencies \cite{Gonchenko2024}. Moreover one can completely characterize the dynamics and structure of the chaotic attractor just by using kneading invariant. However, the discrete Lorenz attractors found in the 3D maps have different properties. Discrete Lorenz attractors are formed by the period-doubling bifurcation of a stable fixed point and then via Neimark-Sacker bifurcation the closed invariant curves they merge. Discrete Lorenz attractors admit homoclinic tangencies and hence are not hyperbolic \cite{Gonchenko2024}. Moreover, since they possess Newhouse regions, their complete bifurcation structure analysis is not possible. Importantly, discrete Lorenz attractors are robust chaotic attractors implying their robustness of chaotic dynamics after perturbations \cite{Gonchenko2021}.  Discrete Lorenz attractors were first found in three dimensional generalized H\'{e}non maps \cite{GONCHENKO2005}. Recently, authors have shown new types of discrete Lorenz attractors after attractor merging crisis of disjoint cyclic Lorenz attractors, coined to be period-two discrete Lorenz attractor \cite{Gonchenko2021}. Discrete maps with axial symmetry posesses many new types of discrete Lorenz attractors such as twisted and non-twisted discrete Lorenz attractors \cite{Gonchenko2024}.           

Networks of dynamical systems is an important aspect to understand the behavior of complex systems \cite{Torres2021}. Such network of coupled dynamical systems display plethora of spatiotemporal patterns varying both in space and time including spiral waves \cite{Ram2024}. Networks of dynamical systems also showcases a novel type of spatiotemporal pattern referred to as a chimera state \cite{Omelchenko2018}. Chimera state refers to the coexistence of synchronous and asynchronous state in a network of identically coupled oscillators. Chimeras are not just a theoretical concept but has been observed experimentally in mechanical systems \cite{Haugland2021}, aquatic animals \cite{Glaze2021} like whale brain which keeps one eye open and other closed. In this present work, we have considered a ring-star network \cite{Muni2020} of 3D sinusoidal map and we illustrate variety of spatiotemporal structures exhibited by the network including traveling waves , synchronized state, incoherence, and chimera state. In this work we have illustrated a variety of spatiotemporal patterns. However, a detailed analysis on the spatiotemporal patterns exhibited by a 3D discrete map showcasing discrete Lorenz attractors remains to be analyzed and will be considered in our next works. 

In the rapidly evolving realm of digital cybersecurity, video encryption stands out as a vital area of research. The main reason being the transmission of sensitive visual datasets across insecure networks \cite{Dhingra2023}. Well known encryption methods such as AES \cite{Devi2019}, DES \cite{Ratnadewi2018} are effective for text based data. However those methods struggle to address the encryption of video data. As an alternative, chaos based encryption methods can be used due to their unpredictable nature and sensitivity to initial conditions. Furthermore, they assure enhanced security, making it relevant for applications in real-time video streaming, military surveillance, IoT security \cite{Francis2024}. In the present work, we have considered a 3D sinusoidal discrete map exhibiting hyperchaotic behavior \cite{MuniHC2024} along with YOLOv10 for object detection and particular object segmentation from the video frame along with CNN (Convolutional Neural Network) generated keys. Each video frame is processed to identify objects of interest and bounding boxes are generated around this object which allows selective encryption instead of the entire image. This approach is efficient and enhances computational efficiency and improves security by focusing on key regions. \textcolor{black}{The video encryption method includes normalizing hyperchaotic sequences followed by pixel substitution, pixel shuffling, and key generation via using feature extraction via CNN to derive cryptographic key from each video frame. The cryptographic key is highly secure and reduces the risk of brute-force attacks. Our algorithm shows promising results on security analysis, differential attack analysis, anomaly detection, robustness to noise analysis as security measures.}

In this paper, we consider dynamics of a three-dimensional sinusoidal map and showcase that the map posesses discrete Lorenz attractors. Due to the sinusoidal terms in the discrete map, the map exhibits the presence of \textcolor{black}{infinitely} many fixed points. They can play an important role in understanding the homoclinic and heteroclinic dynamics of the map.  To understand the maximum number of fixed points exhibited by the 3D map,  a two-parameter color-coded plot is shown colored according to the number of fixed points. It can be observed that the map exhibits the maximum number of five fixed points in that region. We prove that as a particular parameter is increased, the number of fixed points increases to infinity. We show that infinitely many fixed points can exist for the 3D map.  To motivate the existence of discrete Lorenz attractors, we have constructed a two-parameter Lyapunov chart. This revealed diverse topologies of chaotic attractors. Moreover, we discuss the route to formation of a discrete Lorenz attractor. We found that the formation of discrete Lorenz attractor takes the route of the formation of a homoclinic butterfly. First, a stable fixed point under the variation of parameters undergoes a period-doubling bifurcation resulting to period-two points. With further variation of parameters,  the period-two point develops two disjoint cyclic closed invariant curves via the supercritical Neimark-Sacker bifurcation. With further variation of  parameters, we see the development of homoclinic butterfly and slowly after the discrete Lorenz attractor develops. Next, we showcase the coexistence of nested quasiperiodic closed invariant and mode-locked periodic orbit. Finally, we showcase a variety of topologically distinct chaotic attractors. \textcolor{black}{Wavy like chaotic attractors is also observed whose wavy characteristic vary with changes in parameters.}    

The paper is organized as follows. In \S \ref{sec:Formation}, \textcolor{black}{we explore the different formation pathways of the discrete Lorenz attractor.} In \S \ref{sec:ThreeDimension}, we introduce the 3D sinusoidal map and discuss the existence of its simplest invariant sets and their bifurcations and also illustrate novel dynamical behaviors including coexistence of ergodic and resonant torus, plethora of topologically distinct chaotic and hyperchaotic attractors.  In \S \ref{sec:DLA}, we identify the regions of existence of chaotic attractors and discrete Lorenz attractors via the use of two-parameter Lyapunov charts and discuss the pseudohyperbolicity property of the discrete Lorenz attraction via the Light Method for Pseudohyperbolicity. In \S \ref{sec:RINGSTAR}, we illustrate various novel spatiotemporal patterns of the ring-star network.  In \S \ref{sec:videoencryption}, we apply the chaotic signals obtained from the map to perform an efficient video encryption. The paper concludes with conclusion and future directions.

\section{Formation pathways of discrete Lorenz attractors}
\label{sec:Formation}
\textcolor{black}{Discrete homoclinic attractors refer to strange attractors in higher-dimensional maps that typically arise from the homoclinic structure of a saddle fixed point}. Therefore, it is important to understand the topological behavior of the saddle fixed point which plays a major role in the classification of discrete homoclinic attractors. Suppose $O$ refers to the fixed point of a three-dimensional map $f$. Let $\lambda_{1}, \lambda_{2}, \lambda_{3}$ \textcolor{black}{denote} the three eigenvalues of the Jacobian $J$ of the 3D map $f$. The fixed point $O$ can be stable, saddle, or unstable. In the case of a 3D map, if all three eigenvalues are less than one in absolute value, the fixed point is stable. If all eigenvalues are greater than one in absolute value, \textcolor{black}{the fixed point} is unstable. The remaining possible cases are: 

a) The saddle fixed point can have one eigenvalue with  \textcolor{black}{absolute value less than one} (stable directions) and two eigenvalues with \textcolor{black}{absolute value greater than one}  (unstable directions). Such fixed points are of the saddle focus type and they correspond to discrete Shilnikov attractors \cite{Karatetskaia2021}. It has been shown that such attractors show  simple routes to hyperchaos in multidimensional maps \cite{MuniHC2024}. 

b) The saddle fixed point can have two eigenvalues less than one in absolute value (stable directions) and one eigenvalue greater than one in absolute value (unstable directions). Such fixed points are of the saddle-focus type and \textcolor{black}{correspond} to discrete figure-8 spiral attractors.

Discrete Lorenz like attractors are homoclinic attractors that contain only one saddle fixed point with two-dimensional stable manifold and one-dimensional unstable manifold, where the eigenvalues of the saddle fixed point satisfy the following conditions:
a) $0<\lambda_{1}<1, -1<\lambda_{2}<0, \lambda_{3}<-1, |\lambda_{1} \lambda_{2} \lambda_{3}|<1$
b) $ |\lambda_{1}| > |\lambda_{2}|$
c) saddle value $\sigma = |\lambda_{1} \lambda_{3}| > 1$, where $\lambda_{1}, \lambda_{3}$ represent \textcolor{black}{the} stable and unstable eigenvalues \textcolor{black}{closest} to the unit circle. We can observe from the above conditions that saddle fixed point $O$ contains the one-dimensional unstable manifold. Moreover, it is important to note that if condition $ |\lambda_{1}| > |\lambda_{2}|$ is violated, \textcolor{black}{this} might result in the formation of a discrete figure-eight attractor \cite{Gon12}. \textcolor{black}{Based on these} three conditions, we can analyze the geometric properties of the discrete Lorenz attractor. 

\begin{figure*}[!htbp]
%\begin{center}
\centering
\includegraphics[width=0.9\textwidth]{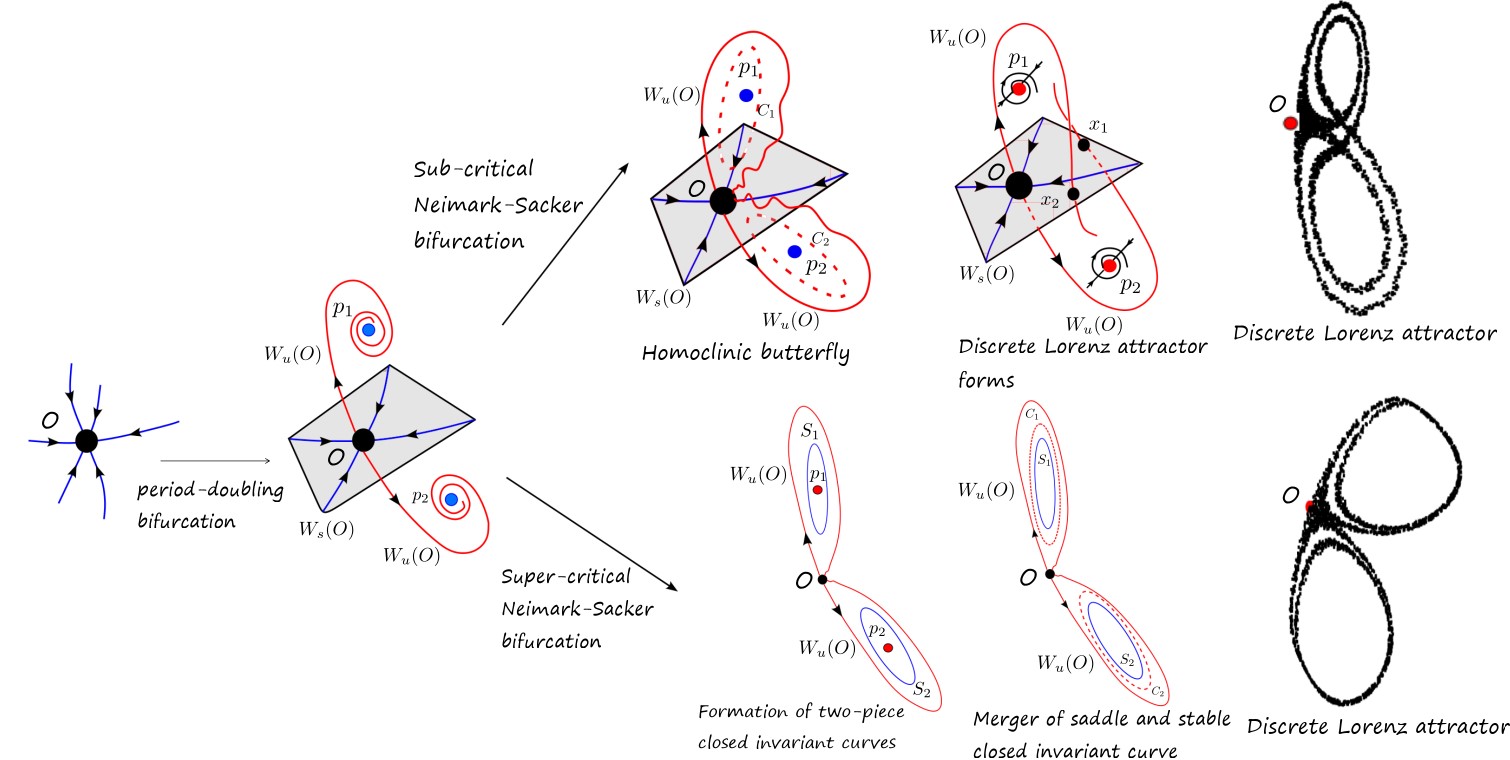}

%\end{center}
\caption{Routes to the formation of discrete Lorenz attractors via a) a subcritical Neimark-Sacker bifurcation and  b) a supercritical Neimark-Sacker bifurcation of a period-two point, \textcolor{black}{which arises from the fixed point $O$ via a period-doubling bifurcation.}}
\label{fig:CollageRoutes}
\end{figure*}

Discrete homoclinic attractors occur in one parameter families of three-dimensional maps \textcolor{black}{through} simple bifurcation routes. The same holds for a special type of discrete homoclinic attractor: the discrete Lorenz attractor. We explain two scenarios leading to the formation of discrete Lorenz attractor. Suppose we have \textcolor{black}{an} asymptotically stable fixed point O. With the variation of a parameter (say $\mu$), the stable fixed point O loses stability \textcolor{black}{through a} period-doubling bifurcation leading to the formation of a period-two orbit $\{ p_{1}, p_{2}\}$ periodic points. The previously stable fixed point O \textcolor{black}{now becomes a saddle type fixed point} with two eigenvalues less than one in absolute value and one eigenvalue greater than one in absolute value. Further variation of parameters leads  to certain bifurcation scenarios: a) The period-two cycle $(p_{1}, p_{2})$ loses stability via a subcritical Neimark-Sacker bifurcation \textcolor{black}{where a} disjoint cyclic closed invariant curve of \textcolor{black}{the} saddle type emerges and further merges with the period-two orbit, see bottom of Fig. \ref{fig:CollageRoutes}. 
b) The period-two cycle $(p_{1},p_{2})$ loses stability via a super-critical Neimark-Sacker bifurcation, leading to the formation of a stable cyclic closed invariant curve, \textcolor{black}{which then merges} with the saddle period-2 cycle and dissappears, see the top part of Fig. \ref{fig:CollageRoutes}.

\begin{figure*}[!htbp]
%\begin{center}
\centering
\includegraphics[width=0.7\textwidth]{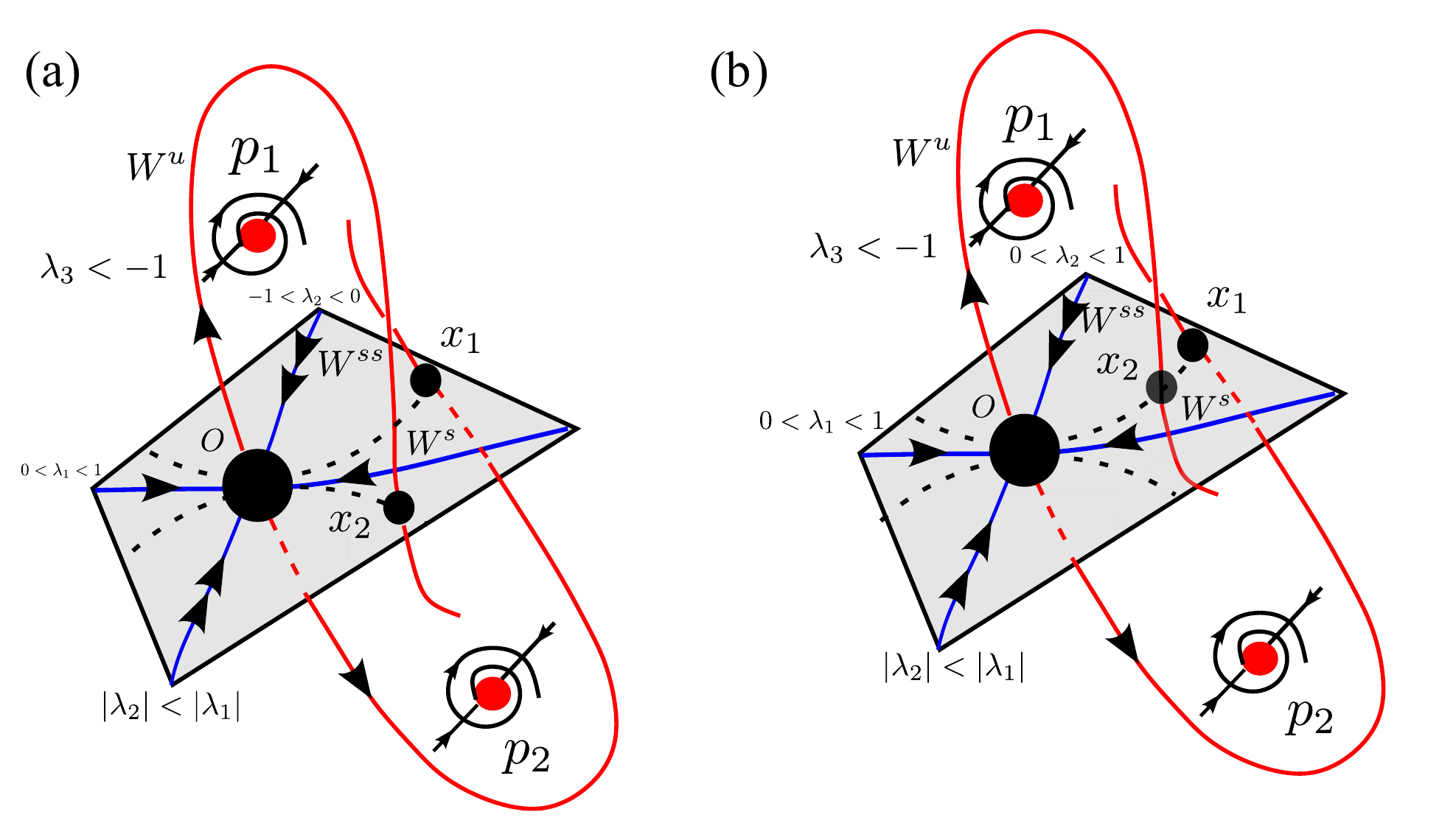}
%\end{center}
\caption{In (a), a schematic of homoclinic butterfly configuration \textcolor{black}{formed} by both branches of the one-dimensional unstable manifold of the saddle fixed point $O$ with one-dimensional unstable manifold and two-dimensional stable manifold with eigenvalues $-1<\lambda_{2}<0<\lambda_{1}<1$ \textcolor{black}{satisfying} $|\lambda_{2}|<|\lambda_{1}|$. Observe \textcolor{black}{that} the homoclinic points $x_{1}, x_{2}$ lie on same side of \textcolor{black}{the} strong stable manifold direction. In (b), a schematic of \textcolor{black}{the} homoclinic butterfly configuration by the one-dimensional unstable manifold of the saddle with eigenvalues $0<\lambda_{2}<\lambda_{1}<1, \lambda_{3}<-1$. }
\label{fig:CollageOrientable}
\end{figure*}

As $\lambda_{3} < -1$, both branches of the one-dimensional unstable manifold emerge from the \textcolor{black}{saddle} fixed point $O$ along the unstable eigen direction of $\lambda_{3} < -1$. \textcolor{black}{Since this} eigenvalue is negative, the points \textcolor{black}{alternate} between the branches. The stable manifold is two-dimensional and consists of \textcolor{black}{both} slow and fast directions, \textcolor{black}{meaning} it has a strong stable manifold. For the case of $-1<\lambda_{2}<0<\lambda_{1}<1$ with $|\lambda_{2}| < |\lambda_{1}|$, the strong stable manifold emerges from $O$, corresponding to the eigen directions \textcolor{black}{associated with the multiplier} $\lambda_{2} < 0$. As a result, the point $O$ on the 2D local stable manifold becomes non-orientable. We can \textcolor{black}{observe} that the strong stable manifold partitions the 2D stable manifold into \textcolor{black}{distinct regions due to $\lambda_{2}>0$}. See Fig. \ref{fig:CollageOrientable} (a), the points cannot jump from one partition to the other partition meaning \textcolor{black}{that iterations are restricted to one side of the partition}. \textcolor{black}{From this we can see that} point $x_{1}$ maps to point $x_{2}$, and both the points lie on only one side of the strong stable manifold. The points $x_{1}, x_{2}$ result from the intersection of unstable and stable manifold. More importantly, the points $x_{1}, x_{2}$ form homoclinic points to fixed point $O$. For non-orientable maps, a similar geometric analogy \textcolor{black}{applies}. Suppose \textcolor{black}{eigenvalues} $\lambda_{1}, \lambda_{2}, \lambda_{3}$ satisfy $0<\lambda_{2}<\lambda_{1}<1$ and $\lambda_{3}<-1$. \textcolor{black}{In this case, we observe that the fixed point $O$ is orientable on the 2D local stable manifold since $\lambda_1 \lambda_2 > 0$}. \textcolor{black}{Consequently,} the homoclinic points $x_{1}, x_{2}$  to $O$ lie on the same invariant curve and don't alternate along the slow manifold, see Fig. \ref{fig:CollageOrientable} (b). \textcolor{black}{An additional observation} from Fig. \ref{fig:CollageOrientable} \textcolor{black}{is related} to the dimensionality of the saddle fixed points $p_{1}, p_{2}$ and $O$. The saddle period-two points $p_{1}$ and $p_{2}$ have two-dimensional unstable manifold and a one-dimensional stable manifold, \textcolor{black}{whereas the} fixed point $O$ has a one-dimensional unstable manifold and a two-dimensional stable manifold. Additionally, the period-two points $p_{1}, p_{2}$ reside in the holes of the discrete Lorenz attractor \textcolor{black}{exhibiting dynamics} similar to the flow like Lorenz attractor.

\section{Three-dimensional sinusoidal map}
\label{sec:ThreeDimension}
The three-dimensional sinusoidal map is \textcolor{black}{defined as follows:}
\begin{equation}
    \begin{aligned}
    x_{n+1} &= y_{n},\\
    y_{n+1} &= sin(z_{n}),\\
    z_{n+1} &= a + bx_{n} + cy_{n} - sin(z_{n}^2),
    \end{aligned}
    \label{eq:STMmap}
\end{equation}
where $x,y,z$ are state variables and $a,b,c$ are parameters. 

It can be shown that 3D map in Eq. \eqref{eq:STMmap} can be transformed \textcolor{black}{into} a third-order difference equation of the form 
\begin{equation}
    z_{n+1} = a + b \sin(z_{n-2}) + c \sin(z_{n-1}) - \sin(z_{n}^2)
    \label{eq:thirdorder}
\end{equation}
For any initial conditions and parameters $a,b,c$, we have 
\begin{equation}
    |z_{n+1}| \leq 1 + |a| + |b| + |c|
\end{equation}
implying that all iterates are bounded and lie inside a cuboid of side $1 + |a| + |b| + |c|$.
\subsection{Fixed points and their bifurcations}
\label{sec:FP}
\begin{figure*}[!htbp]
%\begin{center}
\centering
\includegraphics[width=0.9\textwidth]{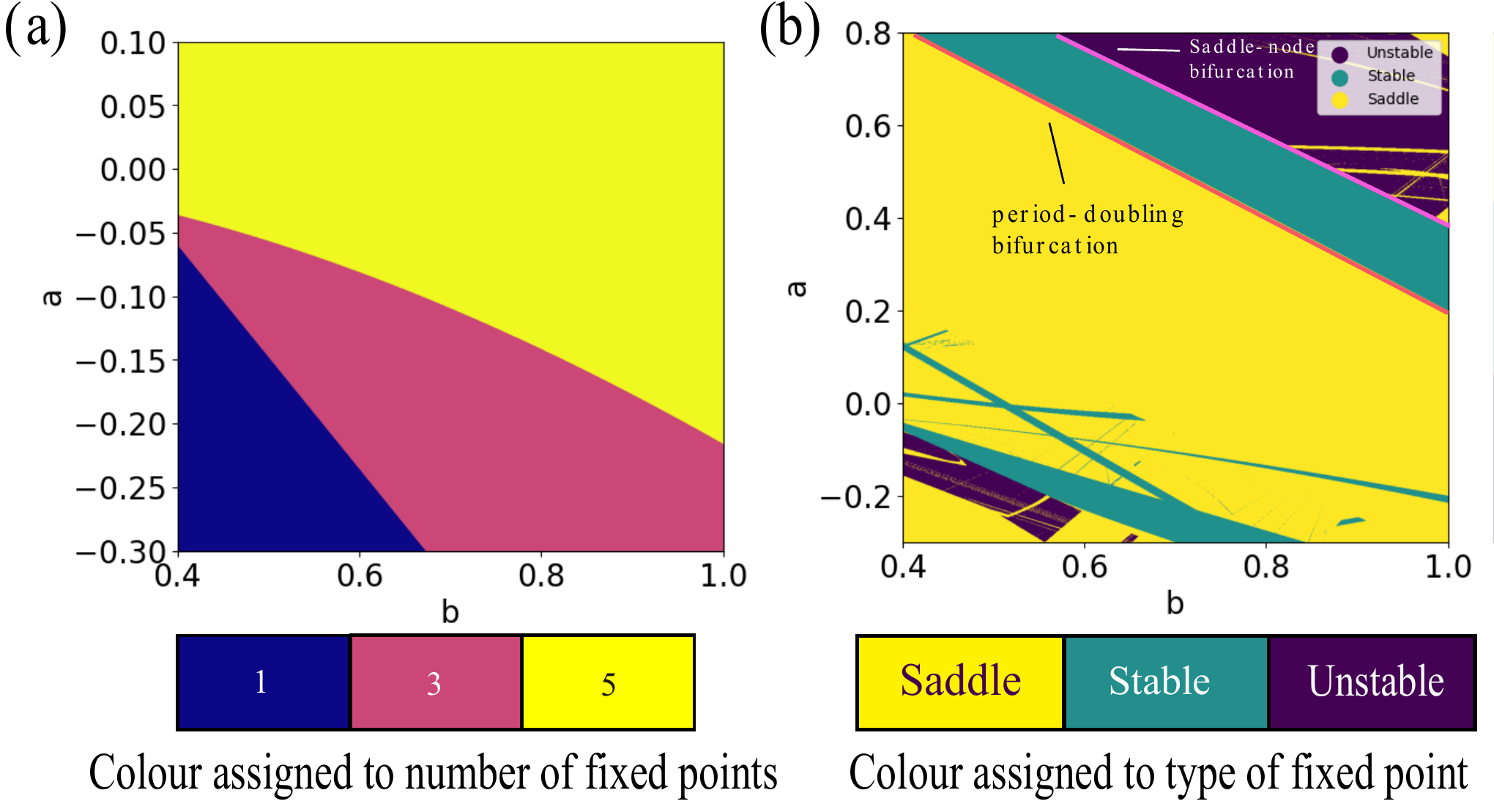}

%\end{center}
\caption{In (a), two-parameter variation of number of fixed points color coded according to the number of existing fixed points.  In (b), a two-parameter stability analysis of fixed points in the $a-b$ plane. The stable fixed point is marked in tile lineated by the codimension-one saddle-node and period-doubling bifurcation curves. The yellow and dark blue region denotes the saddle and unstable fixed point regimes respectively. Parameter $c=0.99$ is fixed.}
\label{fig:CollageStabilityRegion}
\end{figure*}

The fixed points $(x^*,y^*,z^*)$ of Eq. \eqref{eq:STMmap} \textcolor{black}{are obtained} by solving \textcolor{black}{the following} three simultaneous equations: 
\begin{equation}
    \begin{aligned}
    x^* &= y^*,\\
    y^* &= sin(z^*),\\
    z^* &= a + bx^* + cy^* - sin({z^*}^2),
    \end{aligned}
    \label{eq:Fpeqns}
\end{equation}
which leads to finding the roots of the nonlinear equation given by 
\begin{equation}
    z^* = a + (b+c) sin(z^*) - sin({z^*}^2)
    \label{eq:fp}
\end{equation}
Under the simultaneous variations of both parameters $a$ and $b$, we observe that the number of fixed points varies. Moreover, the map \textcolor{black}{exhibits multiple} fixed points ranging from one fixed point to five fixed points, as shown in Fig. \ref{fig:CollageStabilityRegion} (a). We hypothesize that the presence of many fixed points \textcolor{black}{results from} the presence of $\sin$ terms in the discrete map equations. It is interesting to \textcolor{black}{observe} the existence of \textcolor{black}{multiple} fixed points, as they can support the occurrence of homoclinic and heteroclinic dynamics. 
The Jacobian matrix is given by 
\begin{equation}
    J = \begin{bmatrix}
        0 & 1 & 0\\
        0 & 0 & cos(z^*)\\
        b & c & -2zcos({z^*}^2)
    \end{bmatrix}
\end{equation}

 The characteristic polynomial is given by
 \begin{equation*}
     P(\lambda) = \lambda^3 + \lambda^2 (2z^*\cos({z^*}^2)) + \lambda (c \cos(z^*)) + b \cos(z^*) = 0
 \end{equation*}
A saddle-node bifurcation occurs when an eigenvalue reaches $1$, \textcolor{black}{while} a period-doubling bifurcation occurs when an eigenvalue reaches $-1$. Thus, a saddle-node bifurcation occurs when $\lambda = 1$, so $P(1) = 1 + 2z^* \cos({z^*}^2) + c \cos(z^*) + b \cos(z^*) = 0$. Similarly, a period-doubling bifurcation occurs when $\lambda = -1$, so $P(-1) = -1 + 2z^* \cos({z^*}^2) + c \cos(z^*) + b \cos(z^*) = 0$. However, obtaining fixed points \textcolor{black}{analytically is not always} tractable, thus, we can compute the codimension-one bifurcation curves numerically, as shown in Fig. \ref{fig:CollageStabilityRegion} (b). 

\begin{figure*}[!htbp]
%\begin{center}
\centering
\includegraphics[width=0.9\textwidth]{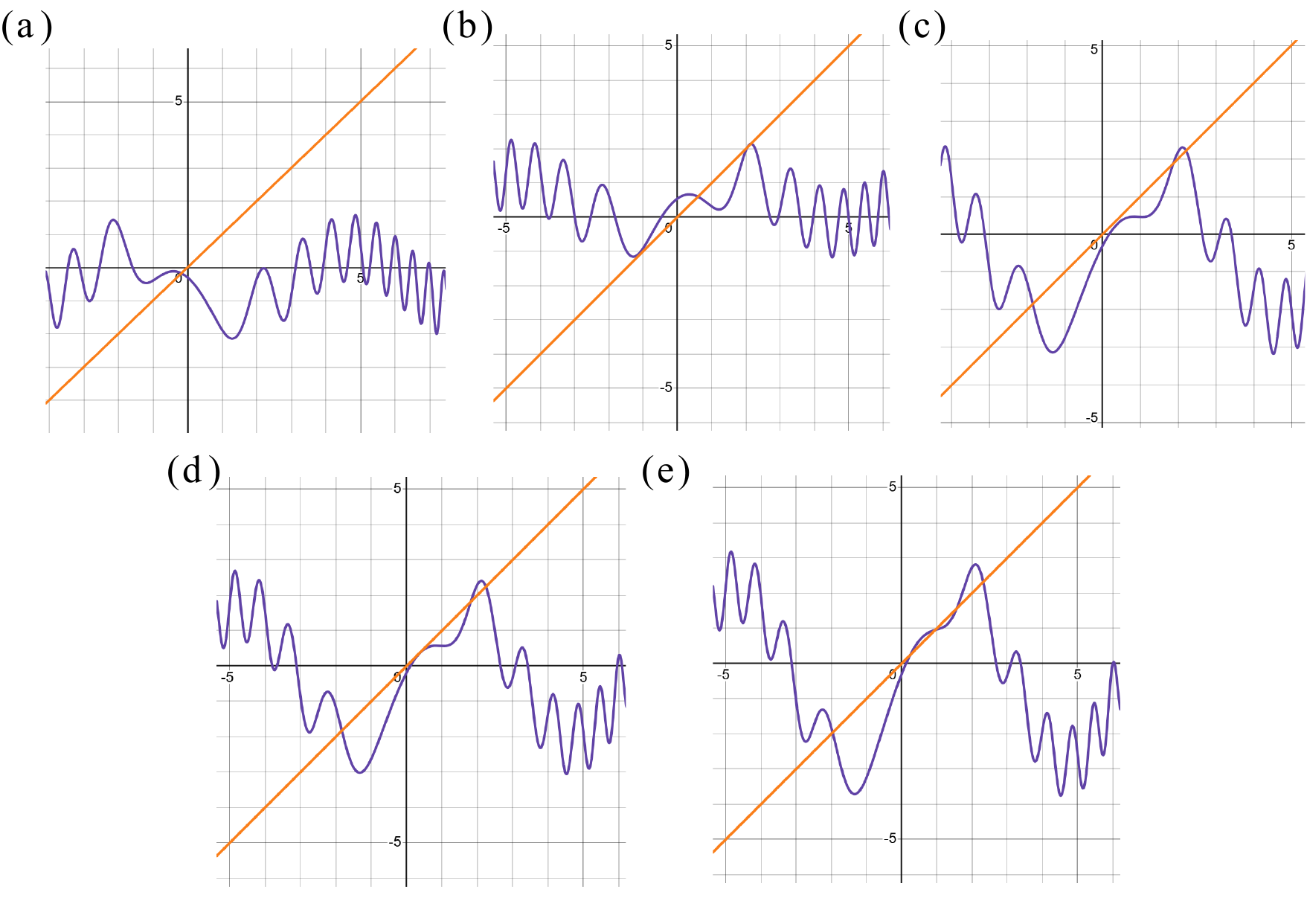}

%\end{center}
\caption{The orange straight line denotes the left side of eq. \eqref{eq:fp} and the blue curve denotes the right 
of eq. \eqref{eq:fp}. In (a), the curves intersect at a single point and thus a single fixed point exists for $a = -0.3, b = 0.416, c = -1.3$. In (b), two fixed points exist at $a=0.54,b=0.725,c=0$. In (c), existence of three fixed points are shown for $a=-0.3,b=0.4,c=1.5$. In (d), existence of four fixed points are shown for $a=-0.19,b=0.4,c=1.5$. In (e), existence of five fixed points are shown for $a=-0.3,b=0.402,c=2.1$.}
\label{fig:NumFixedPoints}
\end{figure*}
To illustrate the existence of \textcolor{black}{multiple fixed points, we use a nullcline-like plot, as seen in} Fig. \ref{fig:NumFixedPoints}. The orange straight line denotes the left hand side of Eq. \eqref{eq:FixedPoint}, while the blue curve denotes the right hand side. The number of intersections \textcolor{black}{between} the orange and blue curve \textcolor{black}{represents} the existence of number of fixed points. In Fig. \ref{fig:NumFixedPoints}, we discuss \textcolor{black}{how the number of fixed points changes depending on various values of  $a,b,c$}. In order to effectively vary all three parameters simultaneously and observe the variation in the number of fixed points, one can think of assigning three parameters $a,b,c$ on three coordinate axis and then scan the parameter space along each quadrant of the parameter axis. In Fig. \ref{fig:NumFixedPoints} (a), we observe one intersection and thus a single fixed point exists for $a =-0.3, b=0.416,c=-1.3$. We observe two fixed points when $a=0.54,b=0.725,c=0$, see Fig. \ref{fig:NumFixedPoints} (b). Existence of three fixed points is noted when $a = -0.3,b=0.4,c=1.5$, see Fig. \ref{fig:NumFixedPoints} (c). With decrease in the value of $a$,  four fixed points emerge at $a = -0.19, b=0.4,c=1.5$, see Fig. \ref{fig:NumFixedPoints} (d). So far, we have observed a maximum of five fixed points for $a=-0.3,b=0.402,c=2.1$, see Fig. \ref{fig:NumFixedPoints} (e). 

However, it is interesting enough that the number of fixed points increases as the parameter $b$ increases. We prove that the 3D map has an infinite number of fixed points as parameter $b$ increases. 

\begin{theorem}
The map contains infinite number of fixed points as parameter $b$ increases.
\end{theorem}

\begin{proof}
To understand the existence of \textcolor{black}{an infinite} number of fixed points in the 3D map, we \textcolor{black}{analyze the number of real solutions to the equation}
\begin{equation}
z^* = a + b \sin(z^*) - \sin({z^*}^2)
\label{eq:FixedPoint}
\end{equation}

Our \textcolor{black}{goal} is to show that the number of real solutions to the above equation is infinite. We \textcolor{black}{achieve} this by \textcolor{black}{demonstrating the existence of at least} $2N$ solutions for arbitrarily large $N$, \textcolor{black}{implying that} the number of solutions can grow \textcolor{black}{indefinitely}. We mention the strategy of the proof: the proof takes advantage of the periodic behavior of $\sin(z)$. The $\sin$ function is periodic and lies in range $[-1,+1]$. Moreover, on any interval  of length $2 \pi$, $\sin(z)$ completes one oscillation. So if any function $f(z)$ also lies within $[-1,+1]$, then the equation $\sin(z) = f(z)$ is guaranteed to have multiple solutions via Intermediate Value Theorem. 

The proof is divided into further steps:
\begin{itemize}
\item \textit{Rewrite the transcendental equation:} 
We can rewrite the equation as $\sin(z) = f(z)$, where $f(z) = \frac{z- |a| -1}{b}$.  The motivation to write in such a way is to split the equation into two parts: a) $\sin(z)$, a well understood oscillatory function, b) $f(z)$, a controllable function whose behavior depends on parameters $a, b$.

\item \textit{Keeping $f(z)$ bounded on $[-1,+1]$:}
For the equation $\sin(z) = f(z)$ to have solutions, $f(z)$ needs to lie between $[-1,+1]$. By choosing parameter $b$ large enough, we can control value of $f(z)$ by making it small (make $b$ large) and ensure $|f(z)|<1$ over a large domain. To perform this, we consider the inequality $\frac{2N\pi - |a| -1}{b} < 1$. This guarantees $f(z)$ is within $[-1,+1]$ for $|z| \leq 2N\pi$, where $N$ is an arbitrary positive integer. Note that the numerator in $f(z)$ grows linearly in $z$ and hence the inequality can be preserved by bounding the numerator. 

\item \textit{Periodic nature of $\sin(z)$:}
Consider an interval of length $2\pi$ for example, $[2n\pi, 2(n+1)\pi]$, the $\sin$ function completes one full oscillation from $-1$ to $+1$. Next, since $f(z)$ is restricted within $[-1,+1]$, there must be at least one solution of $\sin(z) = f(z)$ in each such sub-interval. 

\item \textit{Counting the number of solutions:}
We can divide the domain $[-2N\pi,2N\pi]$ into $2N$ disjoint intervals each of length $2 \pi$ as $[2n\pi,2(n+1)\pi]$, where $-N \leq n \leq N$. As there exists at least one solution in each interval, there are at least $2N$ solutions in total.  

\item \textit{Deducing the count when $N$ is arbitrarily large:}
By allowing $N \rightarrow \infty$, we find that the number of solutions grow without bound. Thus, the number of solutions to the equation is infinite.
\end{itemize}
\end{proof}

\subsection{Coexistence of ergodic and resonant torus}
\label{sec:Coexistence}
\begin{figure*}[!htbp]
%\begin{center}
\centering
\includegraphics[width=1.1\textwidth]{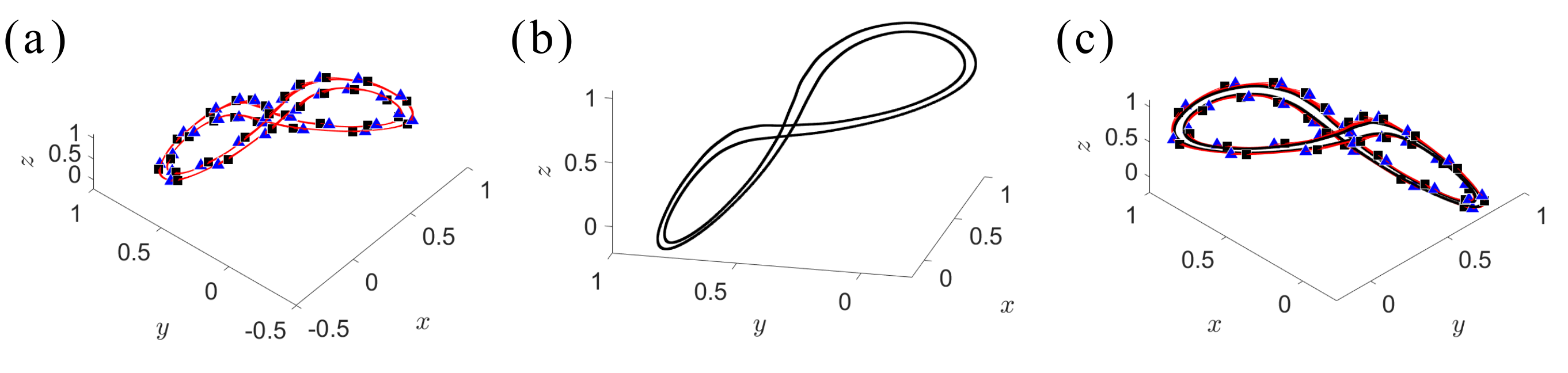}
%\end{center}
\caption{For $a = 0.0192$, we observe coexistence of a length-doubled quasiperiodic closed invariant curve and a mode-locked period-34 orbit. In (a), a mode-locked symmetrical period-34 orbit is displayed. In (b), a length-doubled quasiperiodic closed invariant curve is shown. In (c), both the mode-locked period-34 orbit and the length doubled quasiperiodic closed invariant curve \textcolor{black}{are depicted together}. Observe that \textcolor{black}{these structures are nested within each other. The parameters $b = 0.8, c=0.99$ are fixed.}}
\label{fig:CollageQuasiperiodicSaddleFocus}
\end{figure*}

We illustrate the coexistence of both ergodic and resonant torus in the 3D sinusoidal discrete map. For $a = 0.0192$, we observe the coexistence of length-doubled quasiperiodic orbit and a mode-locked period-34 orbit for different initial conditions; see Fig. \ref{fig:CollageQuasiperiodicSaddleFocus} (c). \textcolor{black}{Interestingly,} the topology of the mode-locked orbit and the quasiperiodic orbit is similar and symmetrical. Moreover, the quasiperiodic orbit and mode-locked orbit are nested to each other. The quasiperiodic orbit orbit \textcolor{black}{resides} inside the mode-locked period-34 orbit. In Fig. \ref{fig:CollageQuasiperiodicSaddleFocus} (a), the symmetrical mode-locked period-34 orbit is \textcolor{black}{depicted, where} stable period-34 orbits are shown in blue triangles, and saddle period-34 orbit are shown as black squares. The corresponding one-dimensional unstable manifolds of the saddle periodic orbit are \textcolor{black}{illustrated} in red \textcolor{black}{forming a} saddle-node connection. A symmetrical length-doubled quasiperiodic orbit with similar topology as that of the mode-locked period-34 orbit is \textcolor{black}{displayed} in Fig. \ref{fig:CollageQuasiperiodicSaddleFocus} (b). \textcolor{black}{Although}, the coexistence of mode-locked and quasiperiodic orbits are common \cite{Muni2022a, MuniHR}, \textcolor{black}{to the best of our knowledge}, such nested mode-locked periodic orbits and quasiperiodic orbits with similar symmetrical  topology are shown for the first time in the \textcolor{black}{context} of 3D maps. 

\subsection{Plethora of chaotic attractors}
\label{sec:Plethora}
\begin{figure*}[!htbp]
%\begin{center}
\centering
\includegraphics[width=0.9\textwidth]{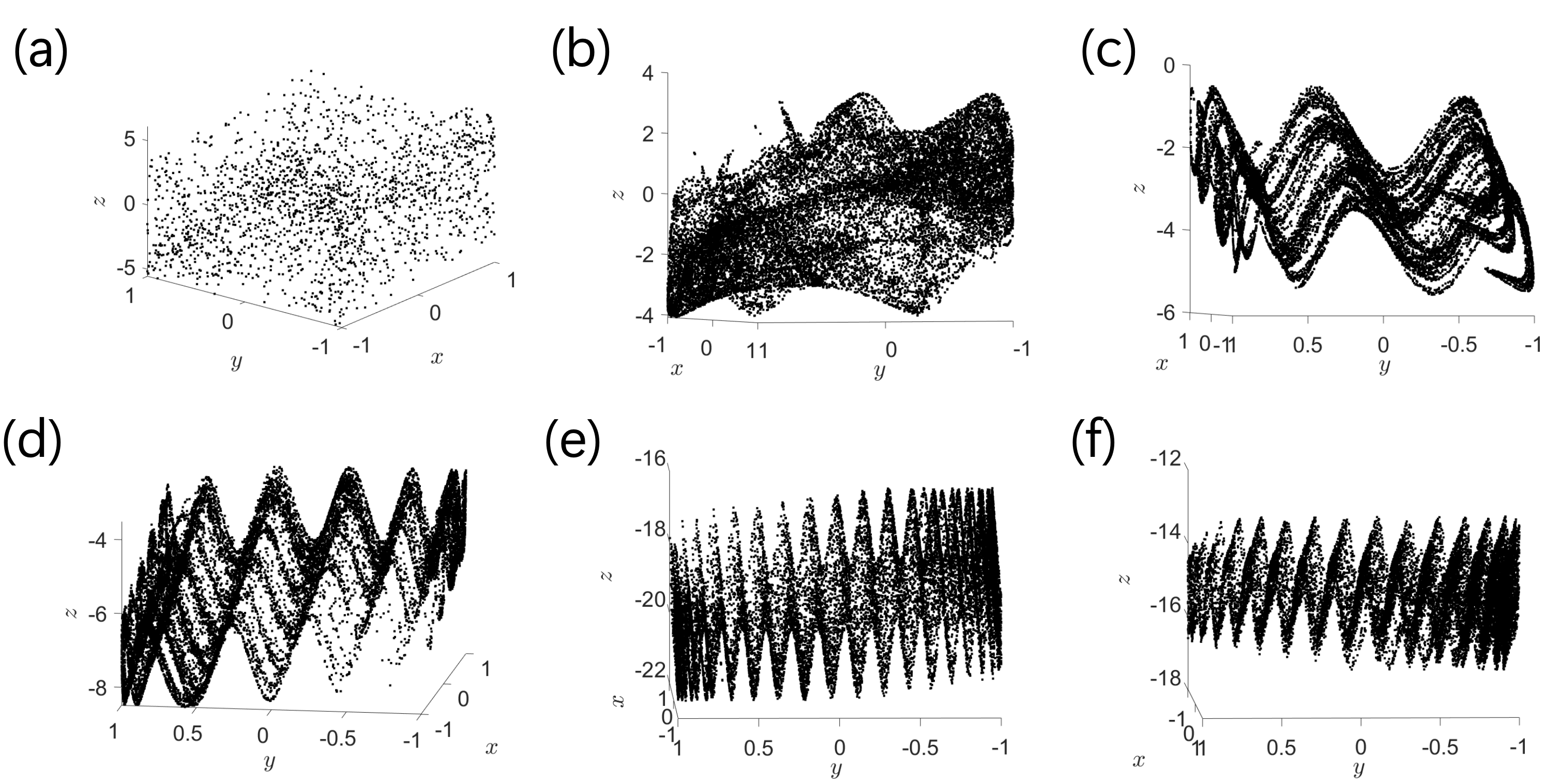}

%\end{center}
\caption{\textcolor{black}{A variety} of chaotic attractors. In (a), we observe a hyperchaotic cube like attractor for $a=0.3, b=3.9,c=0.99$. In (b), a chaotic attractor with different topology for $a=-0.3,b=-1.8,c=0.99$. In (c), a wavy chaotic attractor for $a=-3,b=-0.5,c=0.99$. In (d), more wavy chaotic attractor  for $a=-6.5, b=-0.5, c=0.99$. In (e), extreme wavy chaotic attractor for $a=-19,b=-0.5,c=0.99$. In (f), more wavy chaotic attractor for $a=-15,b=-0.5,c=0.99$.}
\label{fig:CollageChaotic}
\end{figure*}

We also showcase various topologically \textcolor{black}{distinct} chaotic attractors exhibited by the 3D sinusoidal map, see Fig. \ref{fig:CollageChaotic}. For \textcolor{black}{parameter values} $a = 0.3, b=3.9, c = 0.99$, the map develops a cube-like hyperchaotic attractor with two positive Lyapunov exponents. \textcolor{black}{For the parameter values \( a = 0.3 \), \( b = 3.9 \), and \( c = 0.99 \), the system generates a hyperchaotic attractor with a distinctive cube-like shape. The presence of two positive Lyapunov exponents confirms the hyperchaotic nature of the system, as it signifies exponential divergence along two independent directions—one of the key characteristics of hyperchaos.}  

\textcolor{black}{Additionally, the attractor’s Kaplan–Yorke dimension is estimated to be approximately 2, implying that the system’s trajectories are primarily confined to a complex, fractal-like structure embedded within a two-dimensional surface in three-dimensional space. The attractor exhibits a dispersed, cloud-like pattern, indicative of significant dynamical complexity and extreme sensitivity to initial conditions.}  

\textcolor{black}{Unlike conventional Lorenz-type attractors, which typically display a well-defined folded structure, the attractor observed in Fig. 6(a) is more diffuse and cube-shaped, suggesting that the underlying dynamics follow a different organizational pattern.} 
The Lyapunov exponents were estimated numerically via Wolf's algorithm \cite{Wolf1985}. The fractal dimension of the chaotic attractor \textcolor{black}{can be determined via} the formula of Kaplan and Yorke \cite{Kaplan1979} \textcolor{black}{is given by}
\begin{equation}
\rm{D_{KY}} = k + \frac{\sum_{i=1}^{k} \lambda_{i}}{|\lambda_{k+1}|},
\end{equation}
where $k$ is the largest \textcolor{black}{integer} for which the sum of decreasingly ordered Lyapunov exponents $\lambda_{i}$ is still positive. For a \textcolor{black}{variety} of chaotic attractors illustrated in Fig. \ref{fig:CollageChaotic} \textcolor{black}{from} the 3D sinusoidal map, we compute the Kaplan Yorke dimension and \textcolor{black}{confirm} that it \textcolor{black}{aligns} with the expectations, see Table \ref{table1}. Observe Fig. \ref{fig:CollageChaotic}(a), a cubelike chaotic attractor with two positive Lyapunov exponents is shown and the Kaplan Yorke dimension is 2, as per expectations. Chaotic attractors with exactly one positive Lyapunov exponent have fractal dimension between $1$ and $2$. It would be interesting to compare the Kaplan-Yorke dimension with the other fractal dimensions as a future work. 

For negative values of $a$, chaotic attractors with diverse topologies are \textcolor{black}{observed}. However, we have not observed \textcolor{black}{chaotic attractors} for positive values of parameter $a$. For $a=-0.3$, we observe a more dense volume spread of chaotic attractor. As parameter $a$ increases more towards negative value, the chaotic attractor starts taking the shape of a wavy shape. It is interesting to note that the fluctuations of the wavy chaotic attractor increases with increase in the negative value of parameter $a$. For $a=-3$, we observe three wavy peaks in the chaotic attractor, see Fig. \ref{fig:CollageChaotic} (c). For $a=-6.5$, we observe a six peak wavy chaotic attractor, see Fig. \ref{fig:CollageChaotic} (d). For $a=-19$, we observe a heavily fluctuating  wavy peak chaotic attractor, see Fig. \ref{fig:CollageChaotic} (e). 
\begin{table}[h]
\centering
\caption{Lyapunov exponents and Kaplan--Yorke dimension for different chaotic attractors in Fig.~\ref{fig:CollageChaotic}.}
\label{table1}
\begin{tabular}{lll}
\hline
\textbf{Figure} & \textbf{Lyapunov Exponents} & \textbf{Kaplan--Yorke Dimension} \\
\hline
a & (0.880, 0.057, -0.258) & 2 \\
b & (0.5006, -0.1095, -0.4654) & 1.814 \\
c & (1.0692, -0.9633, -1.4456) & 1.066 \\
d & (1.8599, -1.4352, -1.8372) & 1.226 \\
e & (2.9420, -2.0256, -2.3335) & 1.389 \\
f & (2.7148, -1.8680, -2.1545) & 1.390 \\
\hline
\end{tabular}
\end{table}

\subsection{Two-parameter Lyapunov charts}
\label{sec:TwoParamLyap}
\begin{figure*}[!htbp]
%\begin{center}
\centering
\includegraphics[width=0.8\textwidth]{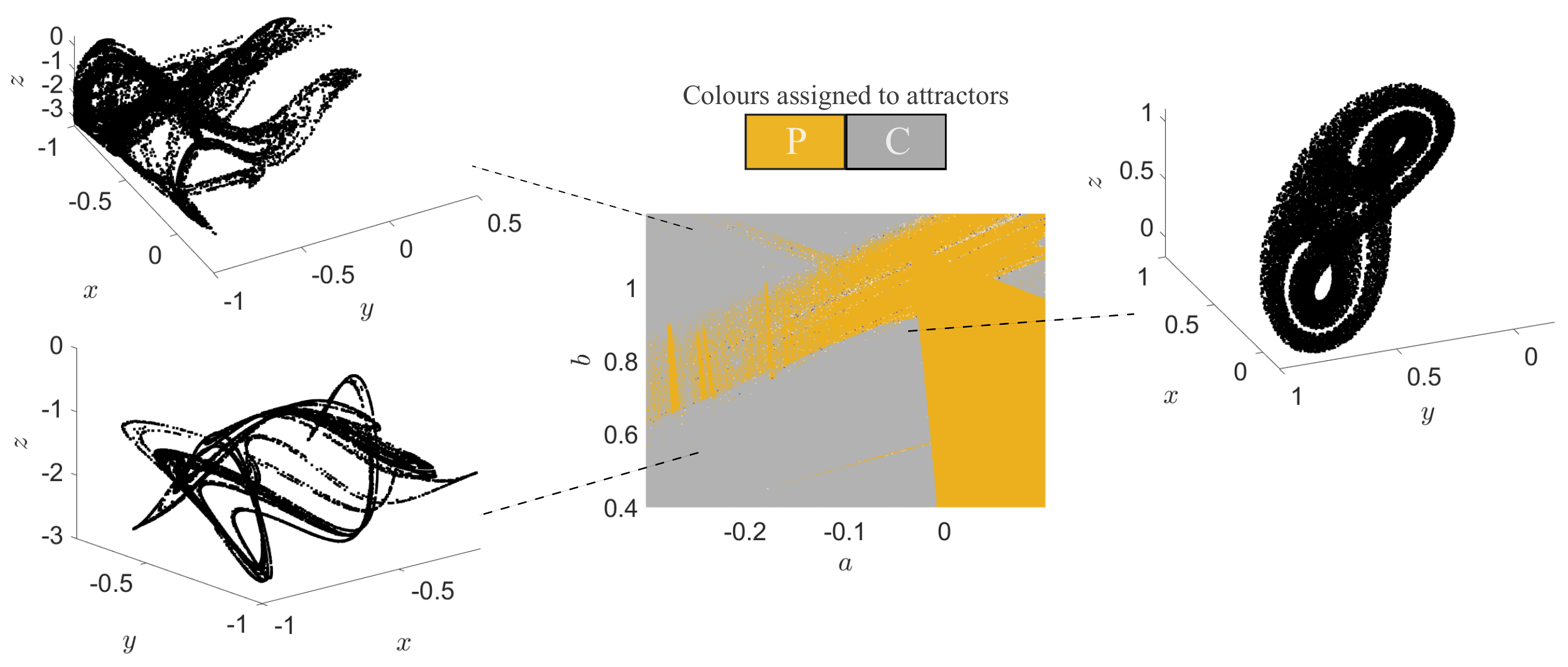}

%\end{center}

\caption{A two-parameter Lyapunov dynamical chart in the $a-b$ plane. The yellow regions denote regular or periodic behavior, \textcolor{black}{while the} grey regions denote chaotic or hyperchaotic behavior. Three different phase portraits are shown in the chaotic regions. We can observe that the topology of the chaotic attractors \textcolor{black}{varies significantly}. One of \textcolor{black}{these} chaotic attractors represent the discrete Lorenz attractor. The parameters $b = 0.8, c=0.99$ are fixed.}
\label{fig:CollageTwoParam}
\end{figure*}

Two-parameter Lyapunov charts are more \textcolor{black}{effective} than one-parameter diagrams in understanding many novel bifurcation scenarios. Two-parameter Lyapunov charts have been useful in understanding the routes to the presence of hyperchaos \cite{MuniHC2024},  routes to the resonant torus doubling bifurcation \cite{MuniTorus}, eyes of chaos \cite{Ramrezvila2024}.
We identify the region of existence of the discrete Lorenz attractors via the use of two-parameter Lyapunov charts in the $a-b$ plane, see Fig. \ref{fig:CollageTwoParam}. The yellow regions denote regular or periodic behavior with all three positive Lyapunov exponents. The grey regions denote chaotic or hyperchaotic behavior with more than one positive Lyapunov exponents. 
For each point of the $1000 \times 1000$ grid in the two-dimensional parameter space, the Lyapunov exponents are computed via Wolf's algorithm \cite{Wolf1985}. \textcolor{black}{If all Lyapunov 
exponents are negative, it denotes the regular or periodic behavior and is marked yellow}. If at least one Lyapunov exponent is positive, it denotes chaotic or hyperchaotic behavior and is marked grey.

\section{Discrete Lorenz attractors}
\label{sec:DLA}

\begin{figure*}[!htbp]
%\begin{center}
\centering
\includegraphics[width=0.9\textwidth]{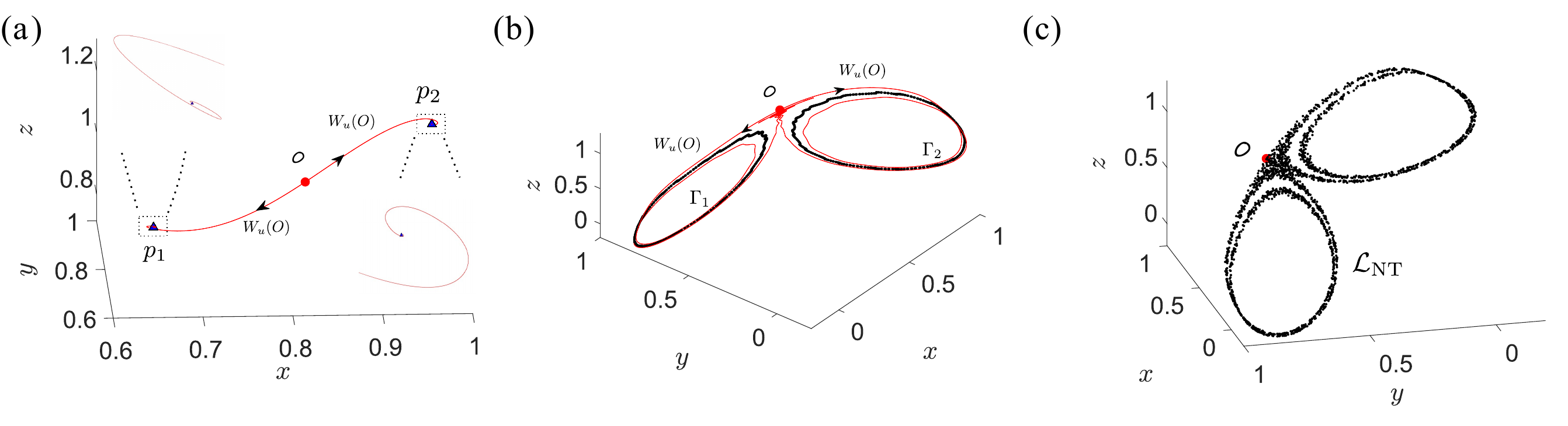}

%\end{center}
\caption{In (a), the fixed point O (shown in red) undergoes a supercritical period-doubling bifurcation 
resulting in a period-two orbit denoted by $\{ p_{1}, p_{2}\}$ at $a=0.3$. Both branches of the one-dimensional unstable manifold are also shown as $W_{u}(O)$ for the saddle fixed point $O$. In (b), the formation of disjoint cyclic closed invariant curves is shown along with the one-dimensional unstable manifolds of the saddle fixed point at $a=0.08$. Observe that the one-dimensional unstable manifolds \textcolor{black}{form a} homoclinic butterfly structure before the formation of the discrete Lorenz attractor.  In (c), the formation of the discrete Lorenz attractor is shown, which has absorbed the saddle fixed point O for $a=0.069$. The parameters $b = 0.8, c=0.99$ are fixed.  }
\label{fig:CollageNonTwistedLorenz}
\end{figure*}

\begin{figure*}[!htbp]
%\begin{center}
\centering
\includegraphics[width=0.9\textwidth]{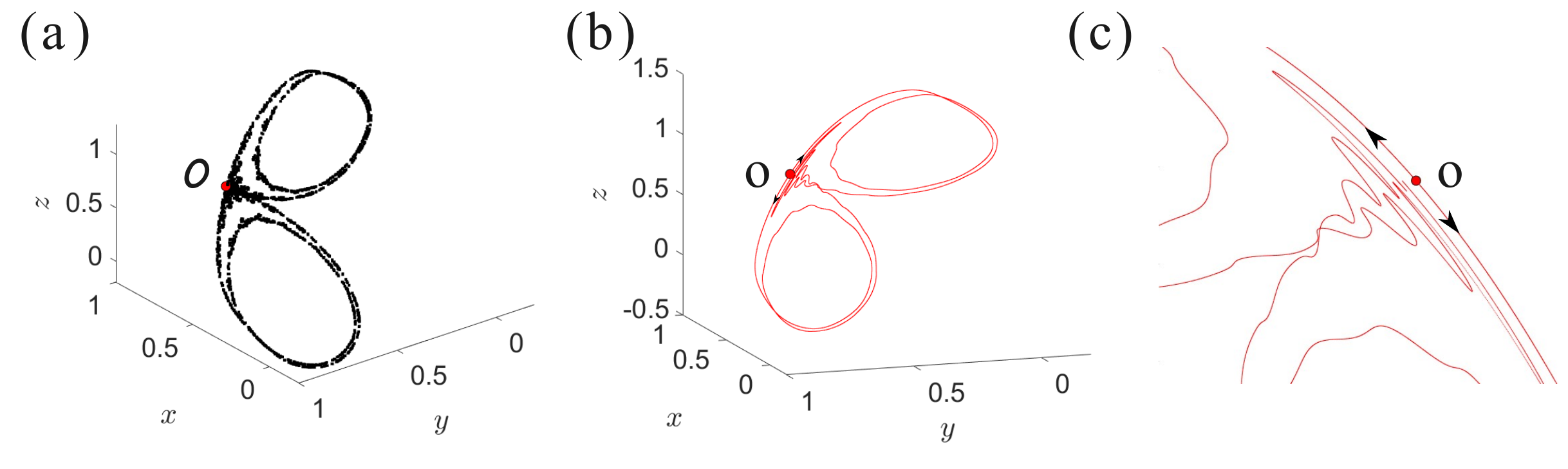}

%\end{center}
\caption{A detailed structure of the discrete Lorenz attractor is shown. In (a), the discrete Lorenz attractor is \textcolor{black}{displayed} along with the saddle fixed point O (in red). In (b), both branches of the one-dimensional unstable manifold of the saddle fixed point O are shown. In (c), a \textcolor{black}{zoomed in view} of the one-dimensional unstable manifold is \textcolor{black}{presented}. We observe the excursions taken by the one-dimensional unstable manifold, \textcolor{black}{where the branches cross each other but do not intersect}.}
\label{fig:DetCollageUnstableManifold}
\end{figure*}

   We \textcolor{black}{demonstrate} the prevalence of discrete Lorenz attractors in the 3D map \textcolor{black}{under consideration}. Importantly, we \textcolor{black}{illustrate} the routes \textcolor{black}{leading} to the formation of a discrete Lorenz attractor by the computation of the one-dimensional unstable manifolds of the saddle fixed point absorbed in the discrete Lorenz attractor. For $a=0.3,b=0.8,c=0.99$,  the stable fixed point O undergoes a period-doubling bifurcation leading to the formation of a stable period-two orbit $\{p_{1}, p_{2}\}$, see Fig. \ref{fig:CollageNonTwistedLorenz} (a). The saddle fixed point  O has one eigenvalue greater than one in absolute value and two eigenvalues less than one in absolute value. Fixed point O develops one-dimensional unstable manifold which converges towards the period-two points spirally shown in a zoomed version.  The period-two point persists as parameter $a$ dereases till $0.08$, after which the stable period-two point undergoes a super-critical Neimark-Sacker bifurcation resulting in the formation of the disjoint cyclic two closed invariant curves. Considering both the branches of the one-dimensional unstable manifold of O, we can see that formation of homoclinic butterfly has taken place. The one-dimensional unstable manifold has grown from the saddle fixed point O covering the closed invariant curve until it returns to the fixed point O where the branches of the unstable manifold appear non-twisted. Further as parameter $a$ decreases to $0.069$, discrete Lorenz attractor develops, see Fig. \ref{fig:CollageNonTwistedLorenz} (c).  
A detailed structure of the discrete Lorenz attractor is shown in Fig. \ref{fig:DetCollageUnstableManifold}. In Fig. \ref{fig:DetCollageUnstableManifold} (a), a detailed structure of discrete Lorenz attractor is shown along with the saddle fixed point O in red, with one-dimensional unstable manifold and two-dimensional stable manifold. In Fig.\ref{fig:DetCollageUnstableManifold} (b), shows only the one-dimensional unstable manifold of the saddle fixed point O forming the homoclinic butterfly. In  Fig.\ref{fig:DetCollageUnstableManifold} (c), a zoomed version of the unstable manifold near by the saddle fixed point O is shown \textcolor{black}{demonstrating} the twistedness of the unstable manifold. 

\begin{figure*}[!htbp]
%\begin{center}
\centering
\includegraphics[width=0.9\textwidth]{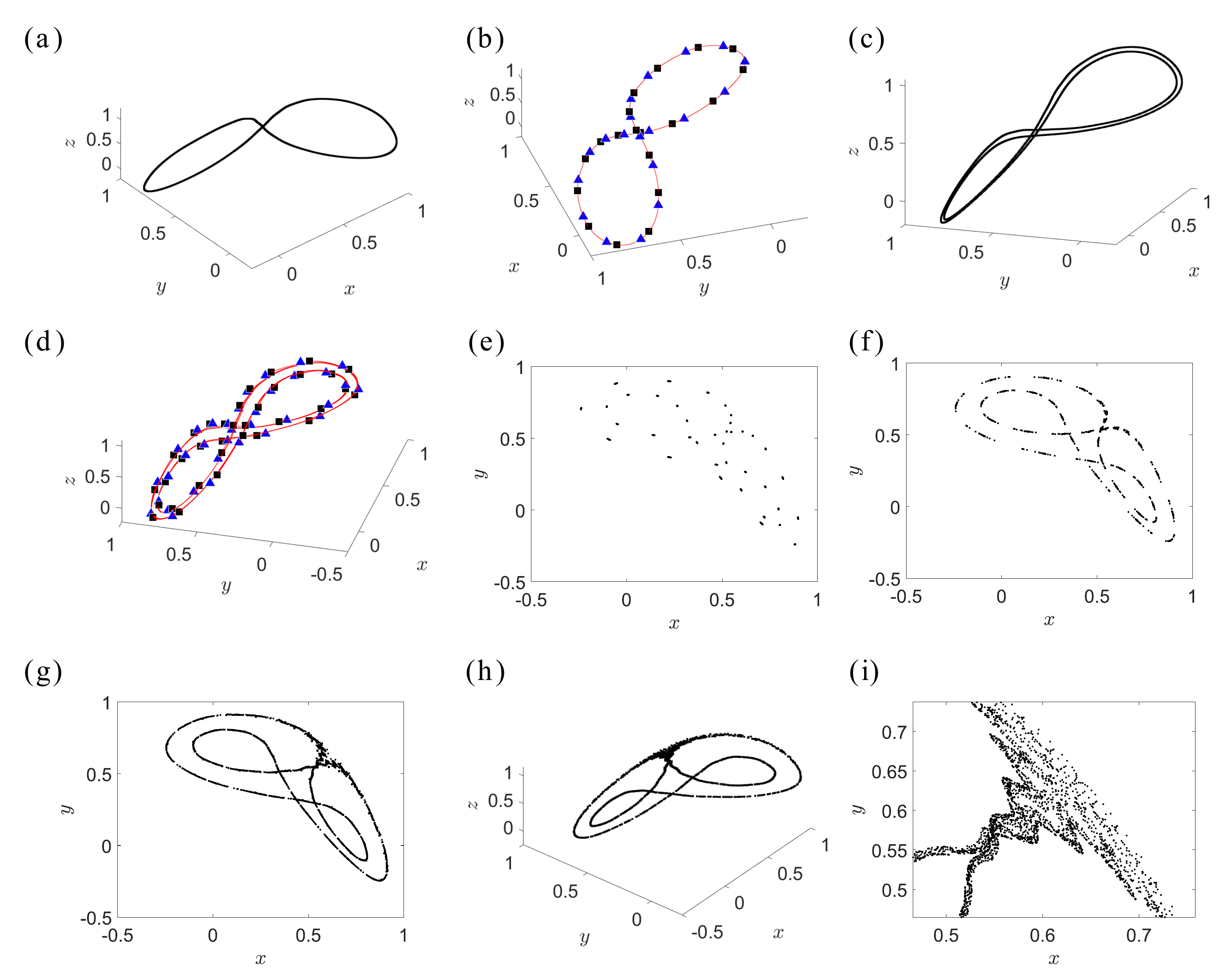}

%\end{center}
\caption{In (a), a quasiperiodic closed symmetric invariant curve is shown for $a = 0.063$. In (b), a symmetric mode-locked period-19 orbit is shown after a saddle-node bifurcation of the invariant curve for $a = 0.062$. In (c), a length-doubled quasiperiodic closed invariant curve is shown for $a = 0.0192$. In (d), a symmetric length-doubled mode-locked orbit of period 34 is shown for $a = 0.019$. In (e), presence of chaotic pieces is shown at $a=0.01527$. In (f), merger of various chaotic pieces is shown at $a=0.015231$. In (g), formation of discrete Lorenz attractor is shown at $a=0.01524$. In (h), fully developed discrete Lorenz attractor is shown at $a=0.015$. The parameters $b = 0.8, c=0.99$ are fixed.}
\label{fig:CollageTwistedLorenz}
\end{figure*}
We show detailed routes \textcolor{black}{leading} to the formation of discrete Lorenz attractor with a different topology in  Fig. \ref{fig:CollageTwistedLorenz}. 
In Fig. \ref{fig:CollageTwistedLorenz} (a), we observe a symmetric closed invariant curve at $a=0.063$. Further variation of parameter $a$ leads to a saddle-node bifurcation of the closed invariant curve \textcolor{black}{resulting in} a mode-locked 
period-19 orbit at $a = 0.062$, see Fig. \ref{fig:CollageTwistedLorenz} (b). The \textcolor{black}{blue triangles} denote the stable period-19 orbit while the black squares denote the saddle period-19 orbit. The red curve \textcolor{black}{represents} both branches of the one-dimensional unstable manifold of the saddles, \textcolor{black}{forming} a saddle-node connection of the mode-locked period-19 orbit. At $a = 0.0192$, formation of asymmetric length-doubled quasiperiodic closed invariant curve takes place, see Fig. \ref{fig:CollageTwistedLorenz}(c). The closed invariant curve is destroyed by the saddle-node bifurcation of invariant curves, leading to the formation of a mode-locked period-34 orbit at $a=0.019$, see Fig. \ref{fig:CollageTwistedLorenz}(d). A further increase in parameter $a$ leads to the formation of chaotic like attractor pieces for $a=0.01527$, see Fig. \ref{fig:CollageTwistedLorenz}(e). The chaotic pieces merge at $a = 0.015231$ \textcolor{black}{taking the} form of a chaotic Lorenz like attractor but it has not formed fully yet, see Fig. \ref{fig:CollageTwistedLorenz}(f). With further variation in parameter $a$ leads to the formation of discrete Lorenz attractor, see Fig. \ref{fig:CollageTwistedLorenz}(g) at $a = 0.01524$.  The discrete Lorenz attractor has fully formed with absorption of the saddle fixed point O at $a = 0.015$, see Fig. \ref{fig:CollageTwistedLorenz}(h). A zoomed version of the attractor is shown, \textcolor{black}{highlighting} the twisted nature of the iterates near the saddle fixed point. 

\begin{figure*}[!htbp]
%\begin{center}
\centering
\includegraphics[width=0.9\textwidth]{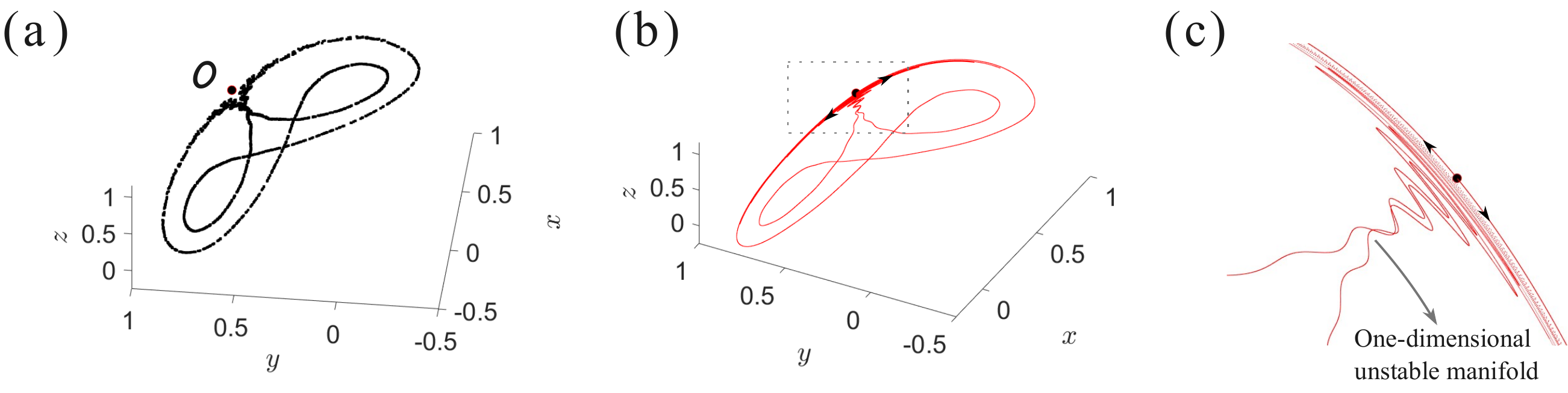}
%\end{center}
\caption{   
 In (a), a twisted discrete Lorenz attractor is shown along with the saddle fixed point O.  In (b), both branches of the one-dimensional unstable manifold  (in red) are shown forming the homoclinic butterfly structure. In (c), a zoomed version of the twisted nature of the discrete Lorenz attractor is shown. The one-dimensional unstable manifold appears to be twisted. }
\label{fig:CollageOneDimensionalUnstableManifold}
\end{figure*}
In Fig. \ref{fig:CollageOneDimensionalUnstableManifold}(a), the twisted Lorenz attractor is shown along with the saddle fixed point $O$. In Fig \ref{fig:CollageOneDimensionalUnstableManifold}(b), both the branches of the one-dimensional unstable manifold of saddle fixed point $O$ is shown in Fig. \ref{fig:CollageOneDimensionalUnstableManifold}(b). We can observe the twisted nature of the one-dimensional unstable manifold. A zoomed version near the saddle fixed point $O$ is shown in Fig. \ref{fig:CollageOneDimensionalUnstableManifold}(c).
\begin{figure*}[!htbp]
%\begin{center}
\centering
\includegraphics[width=0.9\textwidth]{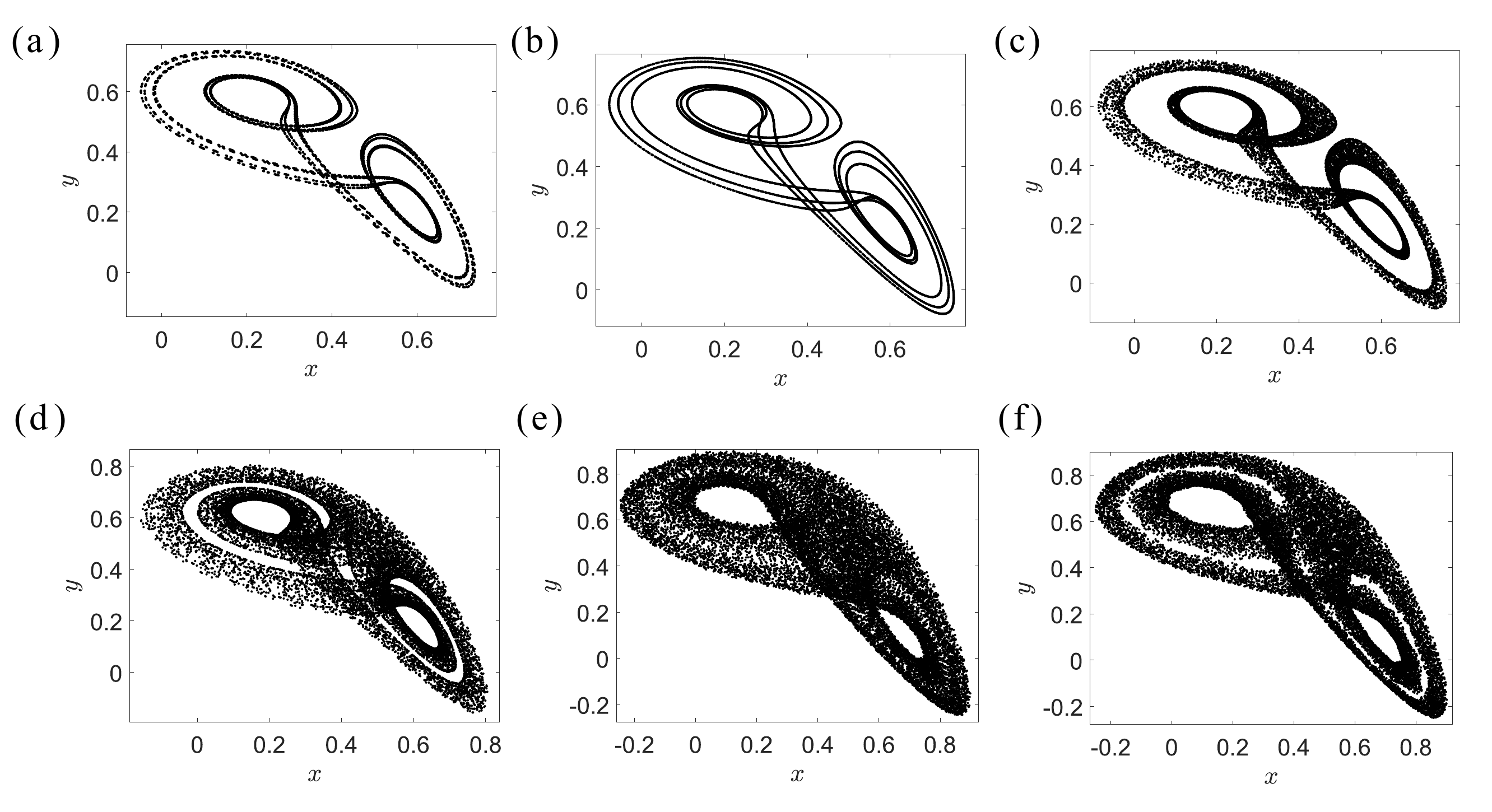}
%\end{center}
\caption{ Formation of a discrete Lorenz attractor for $a<0$. The detailed formation of the discrete Lorenz attractor is shown. It results from the closed invariant curve.}
\label{fig:CollageNegaDLA}
\end{figure*}

So far we have shown the presence of discrete Lorenz attractors in the case of positive values of parameter $a$.    \textcolor{black}{Next, we show} that the discrete Lorenz attractor also exists for negative values of parameter $a$, see Fig. \ref{fig:CollageNegaDLA}. At $a = -0.08$, three layered closed invariant curve exists. Further increase in parameter $a$ to $-0.076$ leads to the formation of a triple length closed invariant curve, see Fig. \ref{fig:CollageNegaDLA} (a), (b). When parameter $a$ is increased to $-0.075$, the closed invariant curve is destroyed and the formation of symmetric chaotic attractor takes place, see Fig. \ref{fig:CollageNegaDLA} (c). At $a = -0.06$, we observe the merger of symmetric heads of the chaotic attractor leading to the formation of a discrete Lorenz attractor, see Fig. \ref{fig:CollageNegaDLA} (d) and it persists for a longer parameter range of $a$. For $a=-0.003$, we observe the merger of wings of the discrete Lorenz attractor  leading to the formation of a two eye discrete Lorenz attractor, see Fig. \ref{fig:CollageNegaDLA} (e). At $a=-0.001$, those wings get detached, see Fig. \ref{fig:CollageNegaDLA} (f) and the discrete Lorenz attractor has similar topology as that of Fig. \ref{fig:CollageNegaDLA} (d). 

\subsection{On verification of pseudohyperbolicity}
\label{sec:LMP}
Pseudohyperbolicity is a generalization of the hyperbolicity property which ensures that the maximal Lyapunov exponent is positive for every orbit of the attractor.  It also guarantees that every orbit in the attractor is unstable. Pseudohyperbolicity is an important aspect to understand the aspect of robustness of chaotic dynamics \cite{Zeraoulia2011}. 
A compact set $\mathcal{A}$ is Pseudohyperbolic if it satisfies the following three conditions:   

a) For each point $x \in \mathcal{A}$, there exists two continuously dependent transversal linear subspaces $\mathcal{N}_{1}(x)$ and $\mathcal{N}_{2}(x)$ with dimensions $\rm{dim}(\mathcal{N}_{1}) = k$ and $\rm{dim}(\mathcal{N}_{2}) = n-k$ for $1 \leq k \leq n-1$. \textcolor{black}{These subspaces are invariant under} the differential of the map $f$ that is $Df(\mathcal{N}_{1}(x)) = \mathcal{N}_{1}(f(x)), Df(\mathcal{N}_{2}(x)) = \mathcal{N}_{2}(f(x))$. In the context of dynamics and hyperbolicity, transversal subspaces usually arise while analyzing the decomposition of the tangent space at points in the attractor. These are associated with splitting of the space into contracting and expanding subspaces. For pseudohyperbolic attractors, such transversality ensures robust instability and further prevents tangencies that could compromise  chaotic dynamics. 

b) The transversal linear subspaces $\mathcal{N}_{1}(x)$ and $\mathcal{N}_{2}(x)$ depend continuously on each point of the attractor $\mathcal{A}$. This continuous dependence ensures that such splitting is well behaved throughout the attractor $\mathcal{A}$ without sudden changes.  To verify the continuity of $\mathcal{N}_{1}(x)$ and $\mathcal{N}_{2}(x)$, we can measure the angle and distance between corresponding points of the attractor. The latter can shed light \textcolor{black}{on whether the} variations are smooth and bounded away from zero \textcolor{black}{helping} to determine the transversality across attractor $\mathcal{A}$.

c) The diferential of the map $Df$ is exponentially contracting in $\mathcal{N}_{1}$ and $Df$ is expanding exponentially in remaining $(n-k)$ dimensions. This means that for \( v \in \mathcal{N}_1 \), the differential \( Df \) satisfies contraction:
\[
\| Df(v) \| \leq C e^{-\lambda} \| v \|, \quad \lambda > 0, \, C > 0.
\]

For \( w \in \mathcal{N}_2 \), the differential \( Df \) satisfies the expansion:
\[
\| Df(w) \| \geq C e^{\sigma} \| w \|, \quad \sigma > 0, \, C > 0.
\]

For a pseudohyperbolic attractor, the fixed point contained \textcolor{black}{within} it must be pseudohyperbolic. That is, the eigenvalues $\lambda_{1}, \lambda_{2}, \lambda_{3}$ should satisfy $|\lambda_{1}|>1, 0< \lambda_{2}<1$, $0<|\lambda_{3}|<|\lambda_{2}|$ and $|\lambda_{1}\lambda_{2}|>1$.
 
For a 3D map, if the attractors are pseudohyperbolic, then ${\rm{dim}(\mathcal{N}_{1})} = 1$ and  ${\rm{dim}(\mathcal{N}_{2})} = 2$.
Three Lyapunov exponents on the attractor $\Lambda_{1}, \Lambda_{2}, \Lambda_{3}$ are determined. They satisfy $\Lambda_{1} >0, \Lambda_{1} + \Lambda_{2} > 0, \Lambda_{1} + \Lambda_{2} + \lambda_{3} < 0$ with $\Lambda_{1} > \Lambda_{2}> \Lambda_{3}$. 

It is not trivial how to implement the above conditions to establish the pseudohyperbolic nature of attractors analytically. A numerical approach to \textcolor{black}{verifying} the pseudohyperbolic nature of attractor was established \textcolor{black}{using the} Light Method of pseudohyperbolicity (LMP method) and was proposed \cite{Gonchenko2018} to study the pseudohyperbolic nature of aattractor in multidimensional systems. Since it is known that for 2d maps for instance, \textcolor{black}{confirming} pseudohyperbolicity is same as confirming uniform hyperbolicity property of the attractor. Then it remains challenging to verify pseudohyperbolicity in multidimensional systems where LMP method can be quite useful. LMP method deals with the most important condition for verifying pseudohyperbolicity: the continuity of subspaces $\mathcal{N}_{1}$ and $\mathcal{N}_{2}$. Using the LMP method, a field of vectors related to the strongest contracting directions is computed. The method \textcolor{black}{involves} in constructing an LMP graph which shows the dependencies of the angles $d\phi$ between the vectors $\mathcal{N}_{1}(x_{1})$ and $\mathcal{N}_{2}(x_{2})$ at orbit points $x_{1}$ and $x_{2}$ of the attractor $\mathcal{A}$. Let us discuss how \textcolor{black}{this method is implemented numerically}: the map is computed for a large number of iterations and transients are discarded. Suppose $p_{i}$ set of points are found. We then compute the corresponding Lyapunov vectors for the map $f^{-1}$. For points $p_{i}$ and $p_{j}$, we compute the distance between them (denoted by $\rho(p_{i},p_{j})$) and angles between them $\phi(p_{i},p_{j})$ and we plot them on a $(\rho,\phi)$ axis. Such dependencies plot will result in a cloud of points in the $(\rho,\phi)$ plane and is referred to as $(\rho,\phi)-$ cloud.  If the $(\rho, \phi)$-cloud has no points other than $(0,0)$ and $(0,\pi)$ on the $\rho=0$ axis, then we can conclude that $\mathcal{A}$ is pseudohyperbolic. Note that the \textcolor{black}{presence} of $(\rho,\phi)$ at $(0,\pi)$ does not affect pseudohyperbolicity. \textcolor{black}{This situation typically arises in} maps whose fixed points, or periodic orbits have negative eigenvalues and also when the attractor itself is non-orientable.
\begin{figure*}[!htbp]
%\begin{center}
\centering
\includegraphics[width=0.7\textwidth]{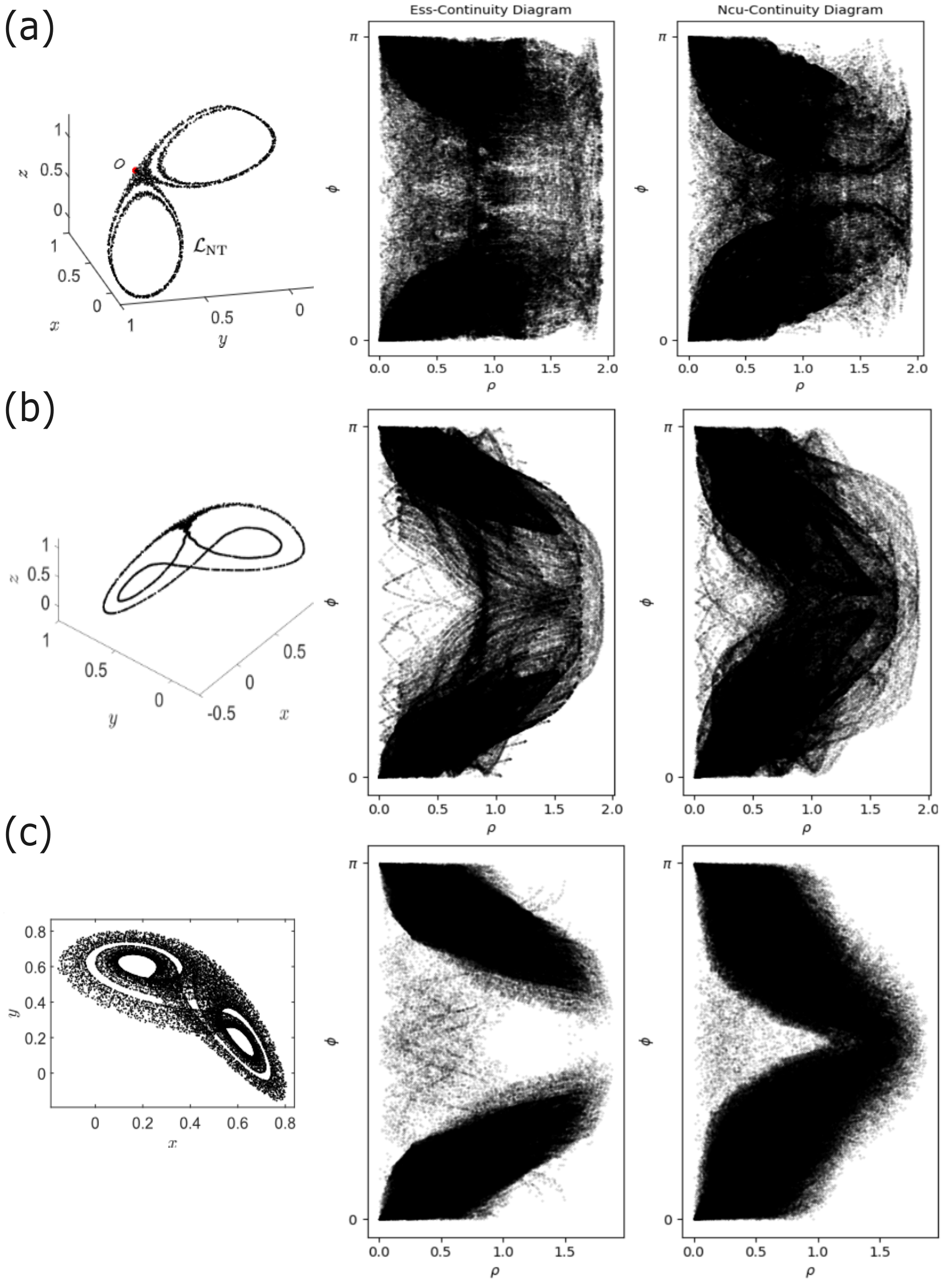}
%\end{center}
\caption{Test of pseudohyperbolicity via the Light Method of Pseudohyperbolicity for three discrete Lorenz attractors. The parameters are fixed as follows: In (a), $a = 0.069$, (b) $a = 0.015$, (c) $a = -0.001$. The parameters $b = 0.8, c=0.99$ is fixed.}
\label{fig:CollagePsuedo}
\end{figure*}
In Fig. \ref{fig:CollagePsuedo}, examples of discrete Lorenz attractors is shown. The continuity diagrams were computed for iterations of the map (discarding the transients). As we can observe, the points touch many points on the $\phi$ axis ($y$-axis) implying the discontinuity of the field of subspaces $E^{ss}(x)$ (subspace spanned by strongly stable direction) and $N^{cu}(x)$ (subspace spanned by strongly contracting direction). This implies the discontinuity of the subspaces and therefore the non-pseudohyperbolicity of the discrete Lorenz attractor.

\section{Ring-star network behaviors}
\label{sec:RINGSTAR}
\begin{figure*}[!htbp]
%\begin{center}
\centering
\includegraphics[width=0.4\textwidth]{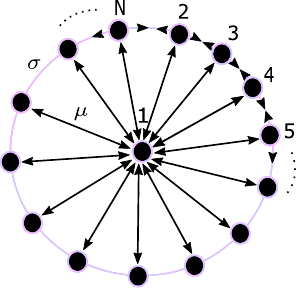}

%\end{center}
\caption{A sketch of ring-star network constituting $N$ nodes with ring coupling strength $\sigma$ and star coupling strength denoted by $\mu$. The double sided arrows indicate bidirectional coupling between the oscillators in the network.}
\label{fig:RingStarNetworkSketch}
\end{figure*}

\textcolor{black}{Researchers have studied phenomena of chaotic synchronization in generalized continuous Lorenz systems \cite{Moon2021}.  
Chaotic synchronization based on Reservoir Computing was performed for the purpose of secure communication\cite{SyncRC}. A novel control scheme for the chaos synchronization in stochastic nonlinear systems was explored \cite{Tan2022}. Higher dimensional chaotic systems were stabilized via optimized linear feedback control and sliding mode control \cite{Laarem2021}.}
A combination of ring and star network result in ring-star network. Moreover, by \textcolor{black}{simply} tuning the coupling strengths of the network, we can switch between a ring network, star network, and ring-star network. Ring-star network was discussed in the \textcolor{black}{context} of Chua circuits where the prevalence of single and double-well chimeras \textcolor{black}{was studied} \cite{Muni2020}. \textcolor{black}{Subsequently}, ring-star network has been \textcolor{black}{applied in various} systems such as neuron systems \cite{Muni2022, MUN2022}, heterogeneous neuron networks \cite{Njitacke2023} \textcolor{black}{among others}. In this network configuration, the network takes the configuration of a ring-star \textcolor{black}{illustrated} in Fig. \ref{fig:RingStarNetworkSketch}.  The ring-star network consists of a central node which connects to other distant nodes of the network. This is beneficial in understanding the information flow throughout the network. In the ring-star sinusoidal map based ring-star network, each node $i = 1, \ldots, N$ \textcolor{black}{follows chaotic dynamics}. The coupling between nodes is bidirectional \textcolor{black}{as indicated by the} double arrows. The coupling strengths of the star and ring networks are denoted by $\mu$ and $\sigma$, respectively. The ring-star network model of the 3D sinusoidal map is described as follows:

\begin{equation}
\begin{aligned}
x_m(n+1) &=  y_m(n) \\
&\quad + \mu (x_m(n) - x_1(n)) + \frac{\sigma}{2R} \sum_{i=m-R}^{m+R}\biggl({x_i(n) - x_m(n)}\biggr), \\
y_m(n+1) &= sin(z_m(n)), \\
z_m(n+1) &= a + bx_m(n) + cy_m(n) - sin(z_m(n)^2), \\
\end{aligned}
\end{equation}

\textcolor{black}{where the central node is defined as}

\begin{equation}
\begin{aligned}
x_1(n+1) &= y_1(n) + \mu \sum_{i=1}^N \biggl( {x_i(n) - x_1(n)} \biggl), \\
y_1(n+1) &= sin(z_1(n))  \\
z_1(n+1) &= a + bx_1(n) + cy_1(n) - sin(z_1(n)^2),  \\
\end{aligned}
\end{equation}

\textcolor{black}{The boundary conditions are \textcolor{black}{given by}}

\begin{equation}
x_{m+N}(n) = x_m(n), \quad y_{m+N}(n) = y_m(n), \quad z_{m+N}(n) = z_m(n),
\end{equation}

where $N$ denotes the total number of nodes in the network, and $R$ \textcolor{black}{represents} the number of neighbors \textcolor{black}{considered for each node}.

Next, we \textcolor{black}{examine} various spatiotemporal patterns exhibited by three different ring-star networks just by tuning the coupling strengths $\mu$ and $\sigma$. For a ring network we can tune $\sigma \neq 0, \mu = 0$. For star network, we can tune $\sigma =0 , \mu \neq 0$. For ring-star network we can tune $\sigma \neq 0, \mu \neq 0$.

\subsection{Ring-star network}
\begin{figure*}[!htbp]
%\begin{center}
\centering
\includegraphics[width=0.6\textwidth]{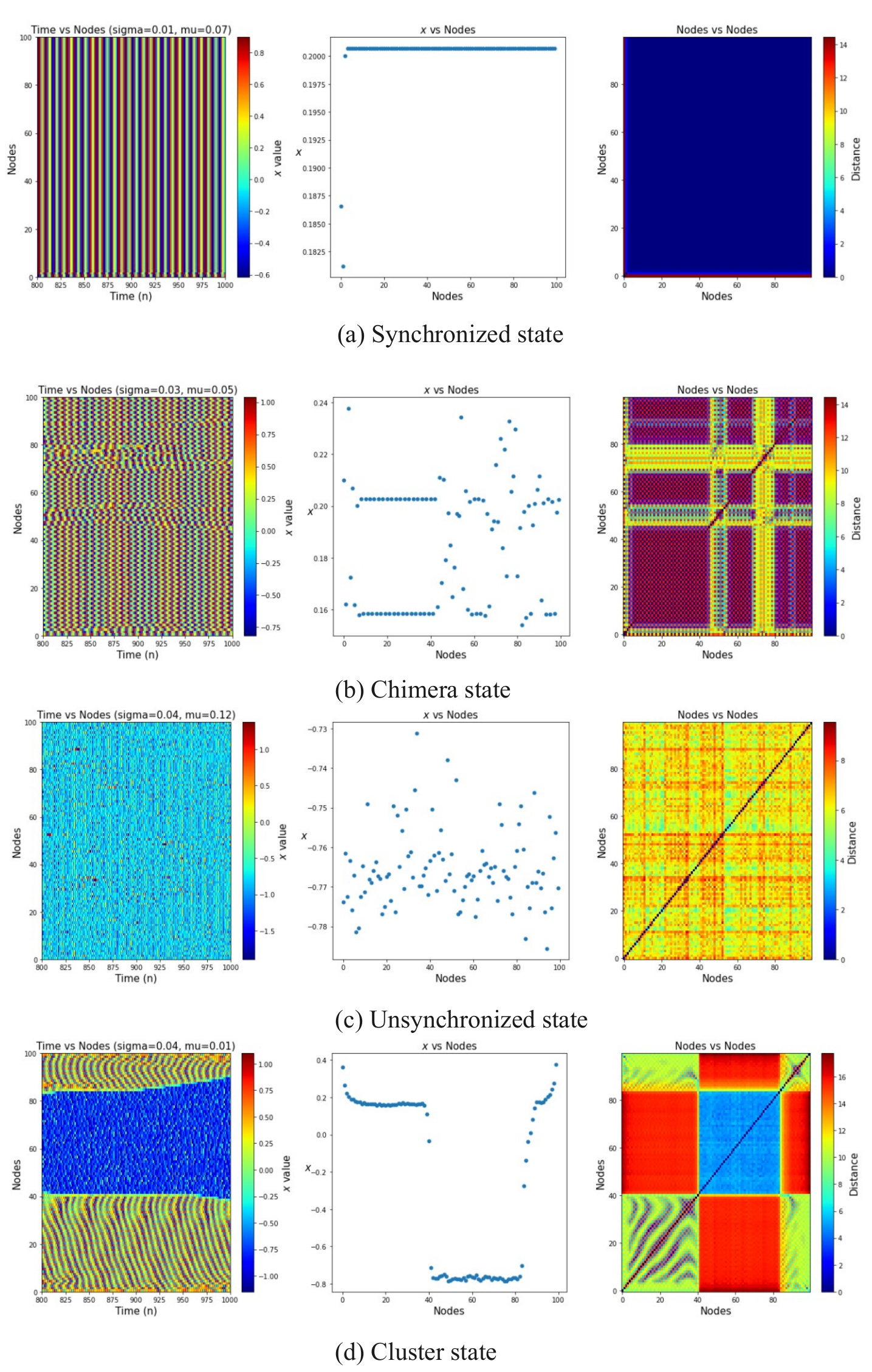}

%\end{center}
\caption{Spatiotemporal patterns exhibited by the ring-star network ($\sigma \neq 0, \mu \neq 0$). In (a), for $\sigma = 0.01, \mu = 0.07$, the network develops a synchronized state. In (b), for $\sigma = 0.03, \mu =0.05$, the network \textcolor{black}{exhibits a} two cluster chimera state. In (c), for $\sigma = 0.04, \mu = 0.12$, the network \textcolor{black}{displays an} unsynchronized state. The dynamical parameters are set as $a = 0.069, b = 0.8, c= 0.99$. }
\label{fig:RingStarNetwork}
\end{figure*}

We consider the ring-star network of the 3D sinusoidal map in the regime of discrete Lorenz attractor, exhibited by uncoupled discrete map at $a=0.069, b=0.8, c=0.99$. For $\sigma = 0.01, \mu = 0.07$, we observe complete synchronization among all the nodes in the ring-star network, see Fig. \ref{fig:RingStarNetwork} (a). The left plot represents the evolution of $x$-state of the nodes with respect to time, \textcolor{black}{where a uniform pattern can be observed}. Corroborating this, the middle \textcolor{black}{plot shows that all oscillators} are aligned at the same level or value showing complete synchronization. The rightmost plot represents the recurrence plot where the distance between the $i$th node and $j$th node is color coded according to the distance. Observe that it is \textcolor{black}{completely blue} \textcolor{black}{indicating that} the distance between nodes is zero and thus denotes complete synchronization. For $\sigma = 0.03, \mu = 0.05$, we observe a two-cluster chimera state, where nodes upto 40 are in a two-cluster state \textcolor{black}{while the remaining nodes} are oscillating asynchronously, see Fig. \ref{fig:RingStarNetwork} (b). The left most plot shows the evolution of the $x$-state over time, while the righmost recurrence plot \textcolor{black}{displays} small square like structures of different colours which denotes the two cluster state. For $\sigma = 0.04, \mu = 0.12$, we observe an unsynchronized state in which all oscillators oscillate out of phase. The evolution of the x-state with respect to time shows no \textcolor{black}{discernible} pattern   
and the recurrence plot \textcolor{black}{reveals} a range of colours, \textcolor{black}{further confirming the unsynchronized state}.

\begin{figure*}[!htbp]
%\begin{center}
\centering
\includegraphics[width=0.5\textwidth]{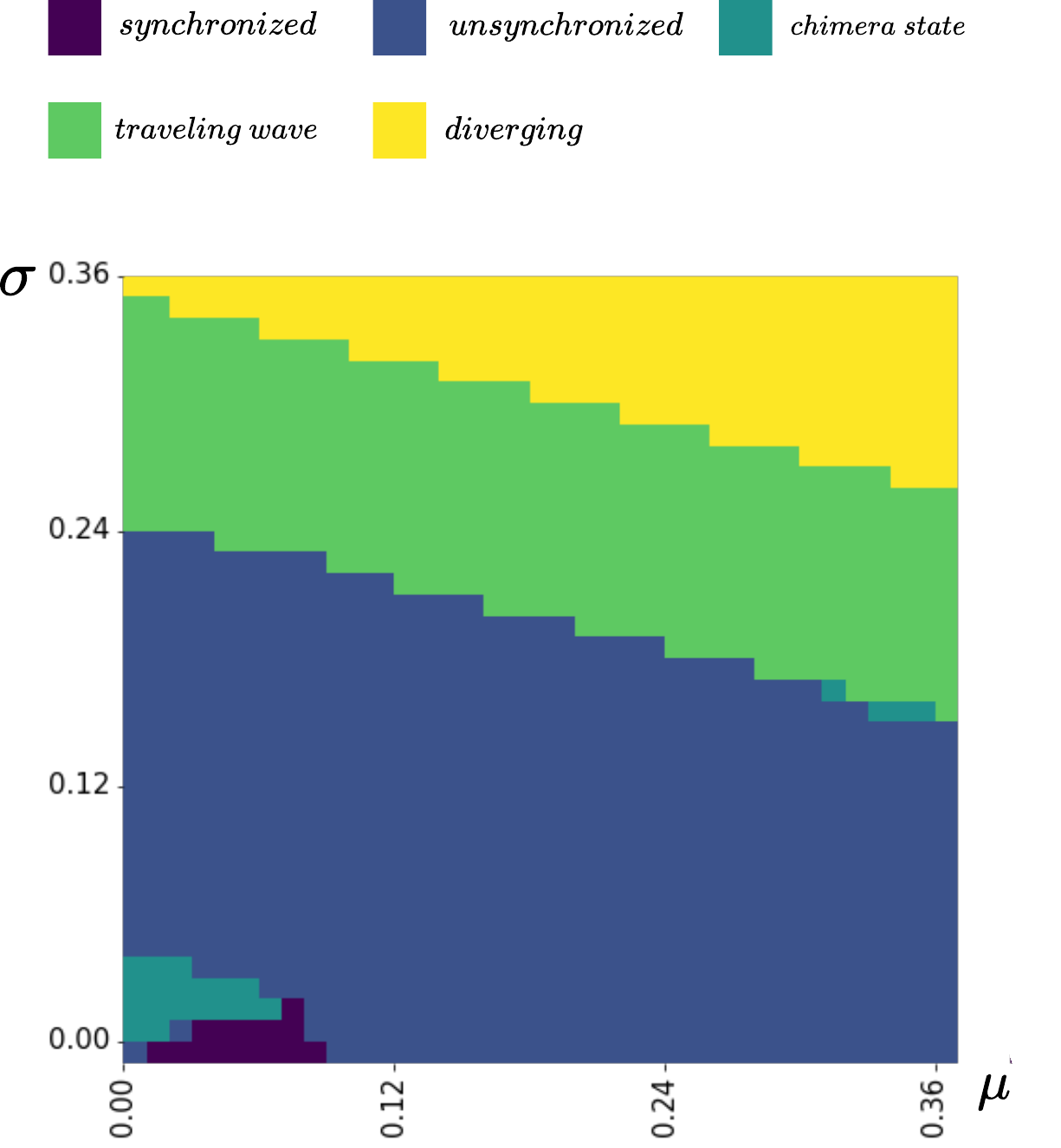}
%\end{center}
\caption{Regime map in the ring and star coupling strength plane. Sensitivity Analysis of Coupling Parameters on Spatiotemporal Patterns.}
\label{fig:RegimeMap}
\end{figure*}

\textcolor{black}{To understand the sensitivity to spatiotemporal patterns with the variations of ring coupling strength $\sigma$ and star coupling strength $\mu$, we perform a two parameter regime map with simultaneous variation of $\sigma$ and $\mu$. Fig. \ref{fig:RegimeMap} illustrates the simultaneous impact of ring coupling strength $\sigma$ and star coupling strength $\mu$ on the emergent spatiotemporal patterns in a ring-star network. The analysis reveals distinct dynamical regimes, where different coupling strengths give rise to synchronized, unsynchronized, chimera, traveling wave, and diverging states. At lower values of $\sigma$ and $\mu$, the system exhibits synchronization and chimera states, indicating partial coherence. As $\sigma$ and $\mu$ increase, unsynchronized and traveling wave patterns emerge, reflecting more complex dynamical behavior. For sufficiently high coupling strengths, the system enters a diverging state, suggesting instability and loss of bounded dynamics. This study highlights the detailed dependence of organization of various spatiotemporal patterns on coupling parameters, offering deeper explanations into the control and predictability of complex networked systems.}

\subsection{Ring network}
\begin{figure*}[!htbp]
%\begin{center}
\centering
\includegraphics[width=0.8\textwidth]{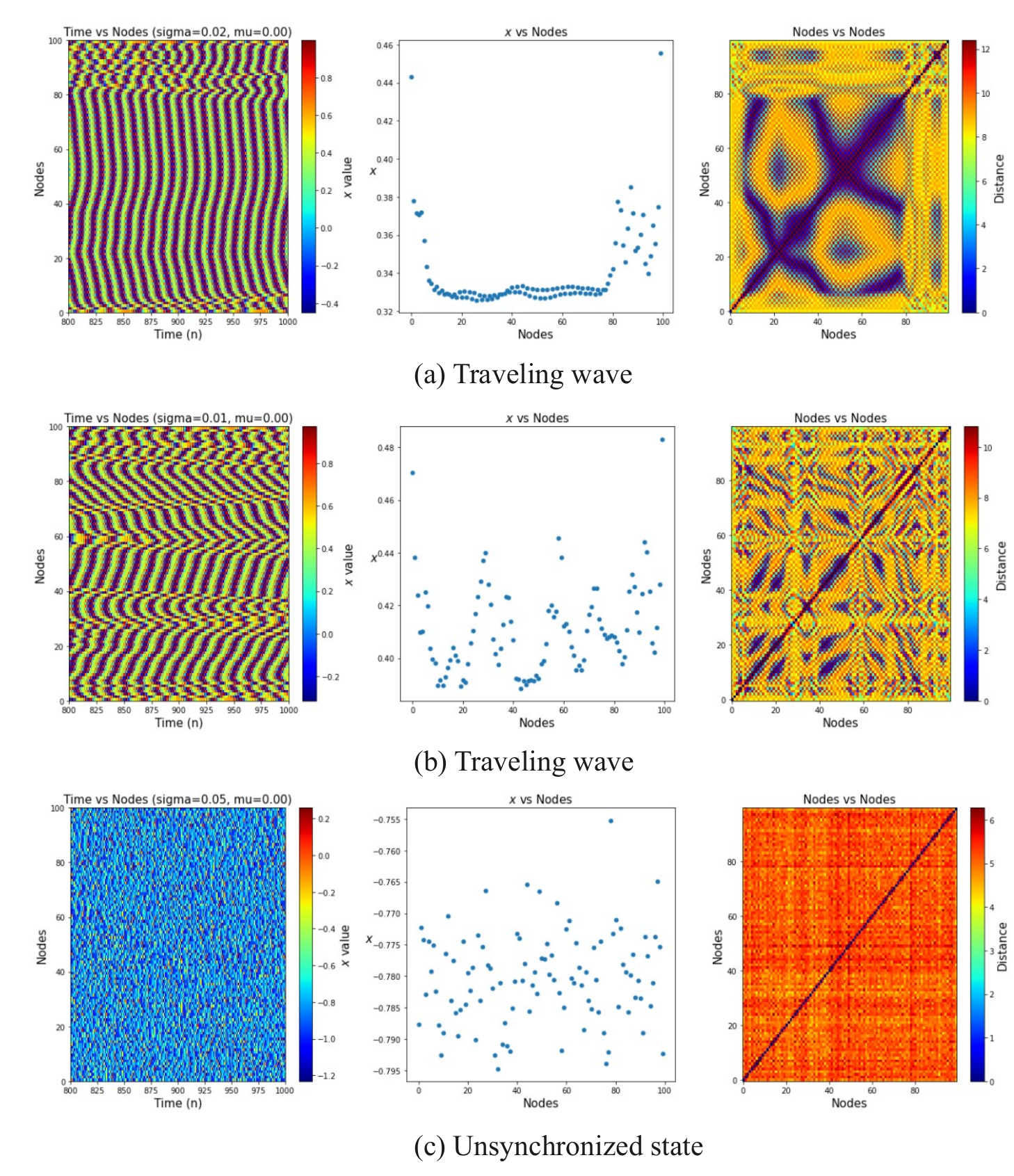}

%\end{center}
\caption{Spatiotemporal patterns exhibited by the ring network. In (a), for $\sigma = 0.02$, the network develops a traveling wave. In (b), for $\sigma = 0.01$, the network develops another traveling wave of different configuration. In (c), for $\sigma = 0.05$, the network develops an unsynchronization behavior. The dynamical parameters are set as $a = 0.069, b = 0.8, c= 0.99$.}
\label{fig:RingNetwork}
\end{figure*}

We consider a ring network of the 3D sinusoidal map by setting $\sigma \neq 0, \mu = 0$. For $\sigma = 0.02, \mu= 0$, we observe a traveling wave, see Fig. \ref{fig:RingNetwork} (a). On the leftmost plot, we can see wavy like patterns in the evolution of the x-state of the nodes of the network over time discarding transients. This coincides with the middle plot \textcolor{black}{which shows} wavy like pattern. In the rightmost recurrence plot, we observe a unique pattern which differs from earlier observed box like structures. This unique structures of the recurrence plots resembles that of the traveling wave state. For $\sigma = 0.01, \mu = 0$, we observe another traveling wave. Notice the similarity \textcolor{black}{between} the evolution of the $x$-state over time in the leftmost plot and also similarity of the recurrence plot on the right. When the ring coupling strength $\sigma$ is increased and tuned to $\sigma = 0.05$, we observe \textcolor{black}{that the nodes oscillate} asynchronously as all oscillators \textcolor{black}{oscillate out} of phase with their neighboring oscillators. 

\subsection{Star network}
\begin{figure*}[!htbp]
%\begin{center}
\centering
\includegraphics[width=0.8\textwidth]{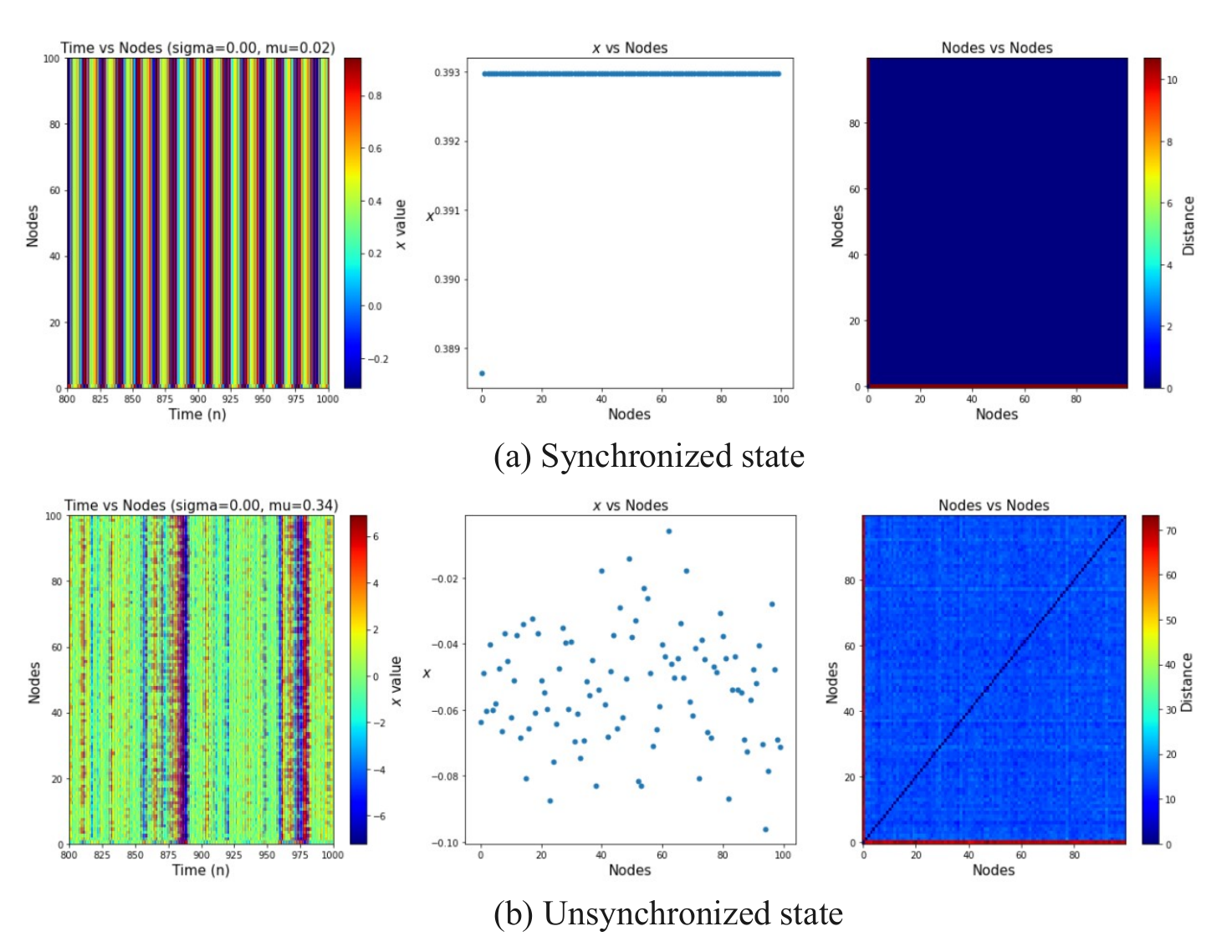}

%\end{center}
\caption{Spatiotemporal patterns exhibited by the star network $(\sigma = 0, \mu \neq 0)$. In (a), for $\mu=0.02$, the network develops synchronized state. In (b), for $\sigma = 0.04$, the network develops unsynchronized state. The dynamical parameters are set as $a = 0.069, b = 0.8, c= 0.99$. }
\label{fig:StarNetwork}
\end{figure*}

For a star network, we set $\sigma = 0, \mu \neq 0$. For $\mu = 0.02$, all oscillators were in sync, see Fig. \ref{fig:StarNetwork} (a). \textcolor{black}{Observe the regularity in the evolution of the $x$-state over time as well as in the rightmost plot which shows a recurrence plot with a complete blue colour indicating synchronization.  With a further increase in $\mu$ to $0.04$, all oscillators were out of sync leading to desynchronization, see Fig. \ref{fig:StarNetwork} (b). This agrees with the evolution of the $x-$state variable over time in the leftmost plot and with the recurrence plot in the rightmost plot.}

\section{Application to video encryption}
\label{sec:videoencryption}
We showcase an application of hyperchaotic signals generated by the 3D sinusoidal map with two positive Lyapunov exponents for the purpose of general multimedia encryption \textcolor{black}{specifically} video encryption. We take advantage of YOLOv10 and CNN to segment the necessary subparts of the video frames for encryption and to generate the secret key respectively. Below each subsection outlines the important steps for video encryption.  

\subsection{Implementation of YOLOv10}
To accurately identify and extract \textcolor{black}{specific} elements from video frames, we employ YOLOv10 for object detection. For our application, YOLOv10 is \textcolor{black}{ideal} because it can detect and locate objects with high precision and real-time performance. Bounding boxes are created around objects of interest that are identified by processing each video frame. By separating target areas, these bounding boxes enable sensitive items to be encrypted selectively rather than the full image. This method focuses on important areas to improve security, decrease computational load, and increase encryption efficiency. High-accuracy real-time processing is made possible by YOLOv10's integration.

\subsection{Encryption using 3D Hyperchaotic Map}

\textcolor{black}{Our motivation for using a 3D discrete map was to highlight its ability to exhibit discrete Lorenz attractors and demonstrate an application in video encryption using hyperchaotic signals derived from the system. While simpler chaotic maps, such as adaptive symmetric chaotic maps \cite{Tutueva2020} and minimal digital chaotic maps \cite{Nepomuceno2019}, may offer similar encryption performance, they often lack the ability to generate discrete Lorenz attractors. This highlights a fundamental trade-off: simpler maps provide lower computational complexity, but they may not capture the intricate dynamics observed in our proposed system.} This paper introduces a novel video encryption method that generates hyperchaotic sequences for pixel transformation and shuffling utilizing signals from a 3D sinusoidal hyperchaotic map.

The encryption process occurs in two stages:
\begin{enumerate}
    \item Pixel intensities are combined with the hyperchaotic sequences to create a transformed frame.

    \item Pixel shuffling is then carried out, using the hyperchaotic map to rearrange pixel locations.

\end{enumerate}

By altering the values and positions of the pixels, this transformation scrambles the image. Without understanding the characteristics of the chaotic system, the combination of these strategies makes it challenging to decipher the encrypted image. Encryption \textcolor{black}{enhances} security and resilience by \textcolor{black}{ensuring} that \textcolor{black}{even minor} modifications to the key or \textcolor{black}{initial} conditions produce \textcolor{black}{significantly} different outcome.

\subsection{ Three Dimensional Hyperchaotic Map}

The encryption algorithm \textcolor{black}{generates} chaotic sequences $x_n$, $y_n$, and $z_n$ for every pixel using a 3D hyperchaotic system \eqref{eq:STMmap}.
 The system is governed by the following equation \eqref{eq:STMmap}, where $x_0$, $y_0$, $z_0$ are the initial conditions and the parameters $a$, $b$, and $c$ govern the dynamics of the system. These sequences are used for pixel transformation and shuffling to ensure high security and robustness.

\subsection*{Normalization of Chaotic Sequences}

The chaotic sequences are normalized in the following \textcolor{black}{manner} to match the image dimensions:

\[
x_{\text{indices}} = x \cdot h, \quad 
y_{\text{indices}} = y \cdot w, \quad 
z_{\text{indices}} = z \cdot 256,
\]
\textcolor{black}{where the height and width of the frame are denoted by $h$ and $w$, respectively}. Here $x_{\text{indices}}$ and $y_{\text{indices}}$ specify the row and column indices, and $z_{\text{indices}}$ modifies the pixel intensity.
\subsection*{ Pixel Substitution (Transformation)}

Each pixel is transformed by adding the corresponding value from $z_{\text{indices}}$, which enhances randomness. The transformation is given by:
\[
\text{Transformed Pixel}[i, j] = (\text{Original Pixel}[i, j] + z_{\text{indices}}) \mod 256,
\]
where $\text{Original Pixel}[i, j]$ \textcolor{black}{represents} the intensity at position $(i, j)$.

\subsection*{ Pixel Shuffling}

Pixels are rearranged based on $x_{\text{indices}}$ and $y_{\text{indices}}$. The new position is determined as:
\[
\text{Shuffled Pixel}[i, j] = \text{Transformed Pixel}[x_{\text{indices}}[i], y_{\text{indices}}[j]].
\]

\subsection*{ Final Encrypted Video Frame Construction}

The encrypted frame is reconstructed by reshaping the shuffled pixels into a 2D format:
\[
\text{Encrypted Image}[i, j] = \text{Shuffled Pixel}[i, j],
\]
where $(i, j)$ represents the position of the pixels in the frame.

\IncMargin{1em}
\begin{algorithm}[htbp]
\subsection{Algorithm : Encryption of Image}
\SetKwInOut{Input}{Input}
\SetKwInOut{Output}{Output}
\SetKwComment{Comment}{\# }{}

\Input{Image $I$ of size $(h, w)$ with pixel values in $[0, 255]$, Key $K$ (string of at least 3 characters)}
\Output{Encrypted Image $E$}
\BlankLine

\textbf{STEP 1: Initialize chaotic system:}\\
Extract the first three characters from key $K$: $K_0, K_1, K_2$.\\
Compute normalized initial conditions: \\
$x_0 = \text{ASCII}(K_0)/255, \quad y_0 = \text{ASCII}(K_1)/255, \quad z_0 = \text{ASCII}(K_2)/255$

\BlankLine
\textbf{STEP 2: Generate chaotic sequences:}\\
Compute the total number of iterations $T = h \cdot w$.\\
Initialize arrays for chaotic sequences $x$, $y$, and $z$ of length $T$.\\
Set $x[0] = x_0, \; y[0] = y_0, \; z[0] = z_0$.\\
\For{$n = 1$ to $T - 1$}{
    $x[n] = a_1 x[n-1] + a_2 y[n-1] + a_3 y[n-1]^2$\\
    $y[n] = b_1 - b_2 z[n-1]$\\
    $z[n] = c \cdot x[n-1]$\\
}
\Comment{Ensure $a_1, a_2, a_3, b_1, b_2, c$ are constants suitable for the map.}

\BlankLine
\textbf{STEP 3: Normalize sequences for indices:}\\
Normalize sequences to valid ranges for indexing: \\
$x_{\text{indices}} = (\lfloor x \cdot h \rfloor) \mod h$\\
$y_{\text{indices}} = (\lfloor y \cdot w \rfloor) \mod w$\\
$z_{\text{values}} = (\lfloor z \cdot 256 \rfloor) \mod 256$

\BlankLine
\textbf{STEP 4: Apply pixel-wise transformation:}\\
Reshape $z_{\text{values}}$ into a 2D array of size $(h, w)$.\\
\ForEach{$i \in [0, h-1], j \in [0, w-1]$}{
    $I'[i, j] = (I[i, j] + z_{\text{values}}[i, j]) \mod 256$
}

\BlankLine
\textbf{STEP 5: Perform pixel shuffling:}\\
Initialize $E$ as a blank image of size $(h, w)$.\\
\ForEach{$i \in [0, h-1], j \in [0, w-1]$}{
    $E[x_{\text{indices}}[i], y_{\text{indices}}[j]] = I'[i, j]$
}

\BlankLine
\textbf{STEP 6: Output encrypted image $E$.}
\end{algorithm}

\subsection*{Key Generation}

\textcolor{black}{The key generation via feature extraction method uses CNN to get a deterministic cryptographic key for each video frame. The CNN produces a feature vector, which is mapped to a fixed-length hexadecimal key to achieve uniformity across frames. The approach focuses on frame-level features, ensuring unpredictability and strength for the encryption. Since keys of CNNs are dynamically computed based on the input data's feature, each example is encrypted distinctively, and therefore the encryption process is safer against brute-force and cryptanalysis attacks through pixel shuffling and linearity of the key. Besides, CNNs are most competent in processing multimedia data, and therefore are most suitable for encrypting images and videos. With the deployment of deep learning, this method enhances security without sacrificing performance, providing a resistant and responsive solution to the challenges of modern-day encryption. Feature extraction highlights critical frame characteristics, strengthening encryption and safeguarding against unauthorized access. Feature extraction is covered in detail in the next section.}

\subsection*{ Feature Extraction and Key Generation}

The following \textcolor{black}{outlines} the procedure for feature extraction and key generation:

\subsection*{ Feature Extraction}
The CNN's convolutional layers process the input image $I$ in order to produce feature maps:

\[
f_1 = \text{ReLU}(W_1 * I + b_1), \quad f_2 = \text{ReLU}(W_2 * f_1 + b_2),
\]
where $W_1, W_2$ are convolutional weights, $b_1, b_2$ are biases, and $f_1$ and $f_2$ are feature maps.

\subsection*{ Feature Vector Creation}
A feature vector $k$ is created by flattening the feature maps and passing them through fully connected layers:

\[
f_3 = \text{ReLU}(W_3 \cdot \text{flatten}(f_2) + b_3), \quad k = W_4 \cdot f_3 + b_4,
\]
where $W_3, W_4$ are weights, $b_3, b_4$ are biases, and $f_3$ is an intermediate feature map.

\subsection*{ Key Normalization and Conversion}
The feature vector $k$ is normalized and converted into a \textcolor{black}{string representation of the key}:
\[
k_{\text{normalized}} = \frac{\text{flatten}(k)}{\max(\text{flatten}(k))},
\]
\[
k_{\text{str}} = \sum_{i=1}^{n} \text{int}(k_i \cdot 255),
\]
where $k_i$ represents each component of the normalized feature vector.

\subsection*{ Hexadecimal Key Generation}
The final cryptographic key is generated by applying SHA-256 hashing to the key string:
\[
k_{\text{final}} = \text{SHA-256}(k_{\text{str}})[0:32],
\]
resulting in a 256-bit hexadecimal key.

\subsection*{ Chaotic Sequence Generation for Weights and Biases}
Chaotic sequences from the logistic map initialize the network's weights and biases:
\[
U_{n+1} = \mu U_n (1 - U_n), \quad V_{n+1} = \mu V_n (1 - V_n),
\]
where $\mu = 3.9999$. After sufficient iterations, weights $W_1, W_2, W_3, W_4$ and biases $b_1, b_2, b_3, b_4$ are sequentially selected from these sequences.

\subsection*{ Decryption using 3D Hyperchaotic Map}

The decryption process accurately reconstructs the original video frame from its encrypted version using the chaotic key stream and neural network-based techniques. Decryption reverses the encryption steps, ensuring precise recovery. The key steps are:

\begin{itemize}
    \item The same chaotic key stream is regenerated using the decryption key, and hyperchaotic signals are subtracted from the encrypted pixel values to \textcolor{black}{restore} the original frame.
    \item Reconstruction is only possible with the correct key, \textcolor{black}{making it essential for decryption}.
\end{itemize}

This approach \textcolor{black}{takes advantage of} the sensitivity and unpredictability of hyperchaotic systems, ensuring strong security and making it highly challenging for unauthorized parties to decrypt the video without the key.

\subsection*{ Decryption Process}

To decrypt the encrypted video, the encryption method based on the chaotic sequence from the 3D hyperchaotic map is reversed. The decryption steps are as follows:

\begin{itemize}
    \item \textbf{Generate Hyperchaotic Sequences:} 
    Using the same initial conditions derived from the key, hyperchaotic sequences $x_n$, $y_n$, and $z_n$ are \textcolor{black}{regenerated} by applying the 3D hyperchaotic map.

    \item \textbf{Normalization of Hyperchaotic Sequences:} 
    The chaotic sequence $z_n$ is normalized to pixel indices as:
    \[
    z_{\text{indices}} = (z_n \cdot 256)
    \]
    \textcolor{black}{This sequence is reshaped into a matrix of size} $h \times w$, where $h$ and $w$ represent the height and width of the encrypted image, respectively.

    \item \textbf{Pixel Transformation (Reversing Encryption):} 
    The chaotic sequence is subtracted from the pixel values of the encrypted image:
    \[
    I_{\text{decrypted}}[i, j] = (I_{\text{encrypted}}[i, j] - z_{\text{indices}}[i, j]) \mod 256
    \]
    where the encrypted image's pixel value is $I_{\text{encrypted}}[i, j]$, and the decrypted image's pixel value is $I_{\text{decrypted}}[i, j]$.

    \item \textbf{Reconstruction of the Original Image:} 
    \textcolor{black}{For RGB images}, this procedure is \textcolor{black}{applied to} every pixel in the image and for every channel.
 The result is a reconstructed image that \textcolor{black}{accurately} matches the original video frame before encryption.
\end{itemize}

\IncMargin{1em}
\begin{algorithm}[htbp]
\subsection{Algorithm : Decryption of Image}
\SetKwInOut{Input}{Input}
\SetKwInOut{Output}{Output}
\SetKwComment{Comment}{\# }{}

\Input{encrypted image, key}
\Output{decrypted image}
\BlankLine

\textbf{STEP 1:} Generate initial conditions from key (ASCII values):\\
$x_0 = \text{ASCII}(k_0), \quad y_0 = \text{ASCII}(k_1), \quad z_0 = \text{ASCII}(k_2)$

\BlankLine
\textbf{STEP 2:} Generate chaotic sequences $x_{n+1}, y_{n+1}, z_{n+1}$ using the chaotic map:\\
$x_{n+1} = a_1 x_n + a_2 y_n + a_3 y_n^2$\\
$y_{n+1} = b_1 - b_2 z_n$\\
$z_{n+1} = c x_n$

\BlankLine
\textbf{STEP 3:} Normalize sequences to pixel indices:\\
$z_{\text{indices}} = \lfloor z_n \cdot 256 \rfloor$

\BlankLine
\textbf{STEP 4:} Reverse pixel transformation:\\
$I'[i, j] = (E[i, j] - z_{\text{indices}}[i, j]) \mod 256$

\BlankLine
\textbf{STEP 5:} Reverse pixel shuffling:\\
$I[i, j] = I'[x_{\text{indices}}^{-1}[i], y_{\text{indices}}^{-1}[j]]$

\BlankLine
\textbf{STEP 6:} Output the decrypted image $I$
\end{algorithm}

\begin{figure}[htbp]
    \centering
    \fontsize{10}{12}\selectfont   %set to 10pt
    \normalsize                    %return to previous font size
    \includegraphics[width=1\linewidth]{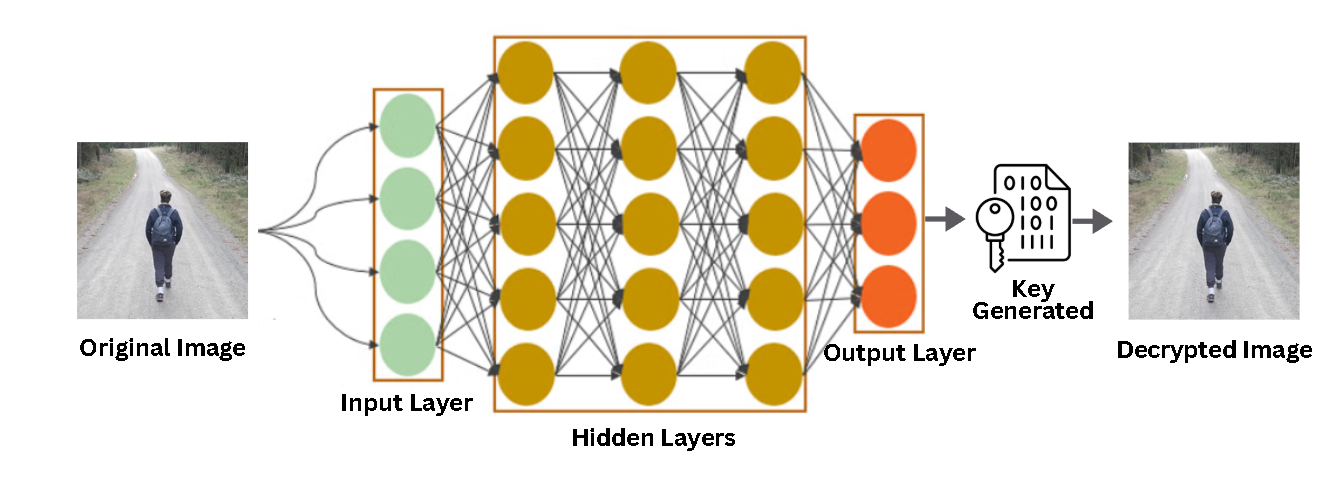}
    \caption{Pathway depicting the process of extracting key information from video frames
for secure encryption and decryption}
    \label{figure:PathwayEncrypted frames}
\end{figure}

\section*{Security Analysis}

The security analysis of video encryption evaluates the effectiveness of encryption techniques in protecting video data from manipulation and unauthorized access. This \textcolor{black}{assessment ensures} the confidentiality, integrity, authenticity, and overall security of video content. 

\begin{itemize}
    \item \textbf{Confidentiality:} 
    Evaluates how well the encryption algorithm \textcolor{black}{prevents} unauthorized decryption attempts, particularly in scenarios where the key \textcolor{black}{remains} unknown.

    \item \textbf{Integrity:} 
    Ensures that the encrypted video remains unaltered during transmission or storage, safeguarding it from tampering.

    \item \textbf{Authenticity:} 
    \textcolor{black}{Verifies} that the encrypted video originates from a legitimate source and has not been compromised or subjected to fraudulent actions.

    \item \textbf{Efficiency:} 
    \textcolor{black}{Focuses on} computational efficiency, ensuring that the encryption and decryption processes do not impose excessive overhead, \textcolor{black}{especially} for large video files or real-time streaming applications.
\end{itemize}

\begin{figure}[htbp]
    \centering
    \fontsize{10}{12}\selectfont   %set to 10pt
    \normalsize                    %return to previous font size
    \includegraphics[width=1\linewidth]{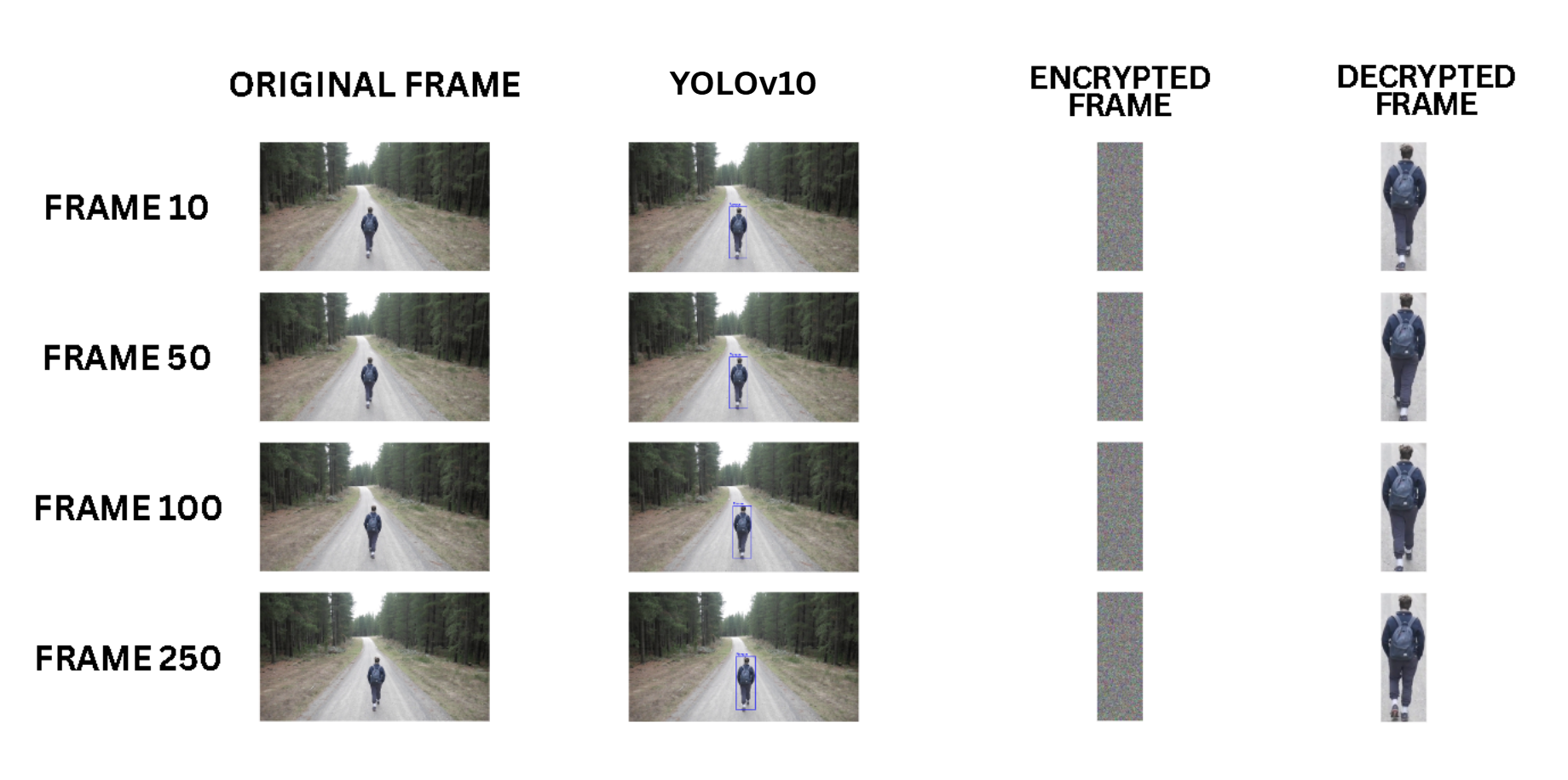}
    \caption{Encryption of frame 10, 50, 100, 250 in the leftmost verticals. In the second column, the image segmentation via YOLOv10 is performed. In the third column, encryption of the segmented part is performed. In the rightmost column, decryption of the encrypted frame  is performed and can be compared with the leftmost vertical as they are identical which shows the efficiency of the video encryption-decryption process.}
    \label{figure:Encrypted frames}
\end{figure}

\subsection*{Differential Attack Analysis}

\textcolor{black}{ Differential attack analysis assesses the strength of an encryption algorithm by examining how small changes in the input affect the output. In this test, the original frame is slightly altered, and the resulting encrypted frame is analyzed to see if it shows any patterns. A strong encryption algorithm should produce entirely different outputs for even minor changes in the input. The frames were encrypted and then decrypted to observe any variations. Metrics like NPCR (Number of Pixels Change Rate) and UACI (Unified Average Changing Intensity) were used to quantify these differences. The results, summarized in the Table\ref{tab:npcr}, show high NPCR values ranging from 99.589 to 99.641 and UACI values from 33.893 to 35.011. These high NPCR values indicate that even slight changes in the input lead to significant variations in the encrypted image, proving the encryption's resistance to differential attacks. The UACI values reflect the intensity changes between the original and decrypted images, further confirming the robustness of the encryption.}
\begin{table}[h]
\centering
\caption{NPCR (Number of Pixels Change Rate) and UACI (Unified Average Changing Intensity) values for encrypted images shown in Figure~\ref{fig:CollageChaotic}.}
\label{tab:npcr}
\begin{tabular}{lll}
\hline
\textbf{Figure} & \textbf{NPCR} & \textbf{UACI} \\
\hline
Figure~\ref{fig:CollageChaotic}(a) & 99.589 & 33.893 \\
Figure~\ref{fig:CollageChaotic}(b) & 99.633 & 34.978 \\
Figure~\ref{fig:CollageChaotic}(c) & 99.637 & 33.938 \\
Figure~\ref{fig:CollageChaotic}(d) & 99.641 & 35.011 \\
Figure~\ref{fig:CollageChaotic}(e) & 99.634 & 33.955 \\
Figure~\ref{fig:CollageChaotic}(f) & 99.603 & 33.893 \\
\hline
\end{tabular}
\end{table}

\begin{table}[h]
\centering
\caption{Comparison of various video encryption approaches along with their correlation analysis, NPCR, and UACI values with the proposed method.}
\label{tab:comparison}
\begin{tabular}{lccccl}
\hline
\textbf{Methods} & \textbf{Horizontal} & \textbf{Vertical} & \textbf{Diagonal} & \textbf{NPCR (\%)} & \textbf{UACI (\%)} \\
\hline
Proposed Method & 0.0019 & -0.0010 & -0.0008 & 99.634 & 33.955 \\
\cite{elkamchouchi2020new} & 0.0025 & 0.0037 & 0.0049 & 98.6327 & 34.0338 \\
\cite{dhingra2024chaos} & -0.0012 & -0.0002 & -0.0006 & 99.65 & 50.02 \\
\cite{sallam2018efficient} & -0.2533 & -0.2557 & -0.2126 & NA & NA \\
\cite{el2023new} & 0.0073 & 0.0052 & 0.0064 & 99.634 & 40.955 \\
\hline
\end{tabular}
\end{table}

\textcolor{black}{\noindent The Table demonstrates the encryption algorithm's performance through NPCR and UACI metrics. High NPCR values (above 99.5\%) and UACI values (near 33\%) indicate strong sensitivity to pixel changes and robust encryption capability.The proposed method in Table\ref{tab:comparison} demonstrates strong encryption performance with a low correlation across horizontal, vertical, and diagonal directions, ensuring minimal statistical dependency in encrypted data. It achieves a high NPCR of 99.634\% and a UACI of 33.955\%, indicating strong resistance to differential attacks. Compared to existing methods, it maintains better unpredictability while offering competitive encryption strength.}

\subsection*{Anomaly Detection}
Anomaly detection evaluates the effectiveness of encryption and decryption by identifying differences between the original and processed frames. \textcolor{black}{Key metrics} such as Mean Squared Error (MSE) and Structural Similarity Index Measure (SSIM) are used. Low MSE values indicate minimal pixel differences, while SSIM values close to 1 demonstrate preserved structural similarity, ensuring encryption robustness and data integrity, see Table \ref{tab:optimized_results} for MSE and SSIM values of various frames.
\begin{table}[h]
\centering
\caption{Performance of anomaly detection for given frames. The Mean Squared Error (MSE) and Structural Similarity Index (SSIM) data demonstrate the accuracy of the anomaly detection methods when applied to different frames of the given video.}
\label{tab:optimized_results}
\begin{tabular}{ccc}
\hline
\textbf{Frame} & \textbf{MSE} & \textbf{SSIM} \\
\hline
Frame 50  & 0.54 & 0.995620 \\
Frame 80  & 0.28 & 0.996712 \\
Frame 100 & 0.45 & 0.997120 \\
Frame 120 & 0.51 & 0.997548 \\
\hline
\end{tabular}
\end{table}

\subsection*{Information Entropy}

Information Entropy measures the uncertainty or randomness of a system. It quantifies how much information is required to represent a system's state. The formula 
\[
H(X) = - \sum P(x_i) \log_2 P(x_i)
\]
calculates entropy, with higher values indicating more unpredictability. In Table \ref{tab:entropy}, the original values represent initial data, while the encrypted values show the data after encryption, introducing uncertainty. The decrypted values indicate the recovered data. The difference between the original and encrypted values highlights the encryption's impact on entropy, while the decrypted values reflect the effectiveness of the decryption process.

\begin{table}[h]
\centering
\caption{Information entropy values for frames at different intervals (50, 80, 100, and 120) in their original, encrypted, and decrypted states. The entropy values of the encrypted frames are consistently close to the ideal value of 8, indicating a high level of randomness introduced by the encryption process.}
\label{tab:entropy}
\begin{tabular}{cccc}
\hline
\textbf{Frame} & \textbf{Original} & \textbf{Encrypted} & \textbf{Decrypted} \\
\hline
50  & 6.89111  & 7.95549  & 6.98696 \\
80  & 6.93351  & 7.95641  & 7.02830 \\
100 & 6.91632  & 7.96185  & 7.01676 \\
120 & 6.98219  & 7.95392  & 7.09219 \\
\hline
\end{tabular}
\end{table}

\subsection*{Robustness to noise analysis}

Robustness to noise is \textcolor{black}{a key factor} in evaluating the strength of an encryption method. This analysis measures how well the encrypted image retains its quality when \textcolor{black}{subjected to} noise, such as compression artifacts or transmission errors, is introduced. Two key metrics—Peak Signal-to-Noise Ratio (PSNR) and Structural Similarity Index (SSIM)—are used to assess the impact. Higher PSNR values indicate less noise interference, while higher SSIM values suggest better preservation of the image’s structure. In the experiments, the encrypted image was subjected to varying noise levels, and the results, as shown in Table \ref{tab:psnr_ssim}, indicate that the encryption technique's performance is impacted by noise, with PSNR and SSIM values remaining low across different clipping ranges. This suggests that the encryption is vulnerable to noise disturbances.

\begin{table}[h]
\centering
\caption{Robustness of the encryption method under varying clip ranges. The table evaluates the impact on PSNR (Peak Signal-to-Noise Ratio) and SSIM (Structural Similarity Index). Each row corresponds to a specific clip range, with consistent PSNR values, indicating minimal degradation in image quality.}
\label{tab:psnr_ssim}
\begin{tabular}{lll}
\hline
\textbf{Clip Range} & \textbf{PSNR (dB)} & \textbf{SSIM} \\
\hline
10--245 & 27.93 & 0.0081 \\
20--235 & 27.93 & 0.0077 \\
30--225 & 27.93 & 0.0071 \\
\hline
\end{tabular}
\end{table}

\subsection*{Performance Comparison of Encryption Methods}
\textcolor{black}{Table\ref{tab:encryptiontime} shows a comparison of encryption times between various approaches, with ``Proposed Method" showing better results. The encryption time, varying from 2.2 to 3.8 over various approaches from related work. The proposed method takes less time that is 2.275, well surpassing other methods, varying from 2.8 to 3.7. This signifies that the proposed method is computationally efficient in encryption processes in comparison to other alternatives. The information highlights the performance of our method, making it a better solution for applications that prioritize optimal encryption performance.}
\begin{table}[h]
\centering
\caption{Comparison of encryption time for different methods. The proposed method demonstrates improved efficiency.}
\label{tab:encryptiontime}
\begin{tabular}{ll}
\hline
\textbf{Encryption Methods} & \textbf{Encryption Time (sec)} \\
\hline
Proposed Method & 2.275 \\
\cite{liang2023new} & 2.881 \\
\cite{Khan2018} & 3.120 \\
\cite{zang2022image} & 3.702 \\
\hline
\end{tabular}
\end{table}

\section{Conclusion}
The paper discussed the prevalence of discrete Lorenz attractors in a three-dimensional discrete sinusoidal map. We \textcolor{black}{examined} the existence of infinitely many fixed points of the map, followed by various codimension-one bifurcation scenarios exhibited by the fixed points. We demonstrated the existence of discrete Lorenz attractors of different topologies. Various routes to the formation of discrete Lorenz attractors were illustrated via phase portraits and two-parameter Lyapunov charts. We also discussed the formation of the butterfly homoclinic structure via the computation of one-dimensional unstable manifold. Furthermore, we explored the spatiotemporal patterns exhibited by the ring-star network of 3D sinusoidal map. We demonstrated the existence of various types of traveling waves, chimera states, and cluster states. Finally, we considered an application of the hyperchaotic signals generated from the 3D sinusoidal map to a general multimedia encryption specifically video encryption. We show that the proposed algorithm efficiently encrypts and decrypts video frames. In general, this work bridges the theoretical understanding of chaotic dynamics with practical applications in cybersecurity, setting the stage for further research on 3D discrete maps, their bifurcations, and innovative real-world implementations. Further classification of discrete Lorenz attractors as orientable or non-orientable remains to be discussed. A detailed unfolding of the spatiotemporal patterns exhibited by the ring network of 3D maps exhibiting discrete Lorenz attractors still remains to be explored which will be considered in our future work.  \textcolor{black}{It would be interesting to explore the dynamics of implicit and semi-implicit maps.} \textcolor{black}{Indeed it is of importance to conduct physical experiments for the verification of theoretical observations obtained in this study. Electronic circuits, microcontroller implementation can be performed to account for experimental setup. Subsequent research can investigate the application of dynamic key management\cite{meena2022secret} to improve the security and performance of the encryption-decryption process. Cryptographic key storage and transmission securely are important to prevent unauthorized access and reduce potential security threats. With the inclusion of strong key distribution mechanisms and encryption protocols, the overall security of the system can be enhanced significantly. In addition, embracing innovative methods like blockchain-based key management\cite{li2021blockchain} or quantum key distribution\cite{cao2022evolution} can also increase the framework's resistance to emerging cyber threats.}

\section*{Acknowledgements}
S.S.M would like to thank Alexey Kazakov for discussions regarding the pseudohyperbolicity of discrete Lorenz attractors. S.S.M also thanks his students Vismaya V S and Anandik N Anand for discussing preliminary results on networks and 
encryption techniques. 
\section*{Conflict of interest}
 The authors declare that they have no conflict of interest.

\section*{Data Availability Statement}
The data that support the findings of this study are available within the article.

\bibliographystyle{unsrt} 
\bibliography{arxiv}
\end{document}